\RequirePackage{fix-cm}
\documentclass[twocolumn]{svjour3}          % twocolumn
\smartqed  % flush right qed marks, e.g. at end of proof
\usepackage{graphicx}
\usepackage{subfigure}
\usepackage{cancel}
\usepackage{amsmath}
\usepackage{amssymb}
\usepackage{dsfont}
\PassOptionsToPackage{hyphens}{url}
\usepackage[colorlinks=true,allcolors=blue]{hyperref}

\begin{document}

\title{A new Bayesian two-sample t-test for effect size estimation under uncertainty based on a two-component Gaussian mixture with known allocations and the region of practical equivalence%\thanks{Grants or other notes
}


\author{Riko Kelter
}


\institute{University of Siegen \at
              Walter-Flex-Strasse 3, 57072 Siegen, Germany \\
              \email{riko.kelter@uni-siegen.de}           \\
              ORCID: 0000-0001-9068-5696
}

\date{\today}

\maketitle

\begin{abstract}
Testing differences between a treatment and control group is common practice in biomedical research like randomized controlled trials (RCT). The standard two-sample t-test relies on null hypothesis significance testing (NHST) via p-values, which has several drawbacks. Bayesian alternatives were recently introduced using the Bayes factor, which has its own limitations. This paper introduces an alternative to current Bayesian two-sample t-tests by interpreting the underlying model as a two-component Gaussian mixture in which the effect size is the quantity of interest, which is most relevant in clinical research. Unlike p-values or the Bayes factor, the proposed method focusses on estimation under uncertainty instead of explicit hypothesis testing. Therefore, via a Gibbs sampler the posterior of the effect size is produced, which is used subsequently for either estimation under uncertainty or explicit hypothesis testing based on the region of practical equivalence (ROPE). An illustrative example, theoretical results and a simulation study show the usefulness of the proposed method, and the test is made available in the R package \texttt{bayest}.
\end{abstract}

\section{Introduction}
\label{intro}
In medical research, the \textit{t}-test is one of the most popular statistical procedures conducted. In randomized controlled trials (RCT), the goal often is to test the efficacy of new treatments or drugs and find out the size of an effect. Usually, a treatment and control group are used, and differences in a response variable like blood pressure or cholesterol level between both groups are observed. The gold standard for deciding if the new treatment or drug is more effective than the status quo treatment or drug is the p-value, which is the probability, under the null hypothesis $H_0$, of obtaining a difference equal to or more extreme than what was actually observed. The dominance of p-values when comparing two groups in medical (and other) research is overwhelming \cite{Kelter2020BayesianPosteriorIndices}, \cite{Kelter2020}, \cite{Nuijten2016}.

The original two-sample t-test belongs to the class of frequentist solutions. These are based on sampling statistics, which allow to reject the null hypothesis via the use of p-values. The misuse and drawbacks of p-values in medical research have been detailed in a variety of papers, including an official ASA statement in 2016 \cite{wasserstein2016}. On the other side, Bayesian versions of the two-sample t-test have become more popular recently. Examples include the proposals in \cite{Gonen2005}, \cite{Rouder2009}, \cite{Wetzels2011}, \cite{Wang2016} and \cite{Gronau}. All of these focus on the Bayes factor (BF) for testing a null hypothesis $H_0:\delta=0$ of no effect against a one- or two-sided alternative $H_1:\delta >0$, $H_1:\delta <0$ or $H_1:\delta \neq 0$. Bayes factors themselves are also not without problems: (1) Bayes factors are sensible to prior modeling \cite{Kamary2014}; (2) Bayes factors require the researcher to calculate marginal likelihoods, the calculations of which can be complex except when conjugate distributions exist; (3) In the setting of the two-sample t-test, Bayes factors weight the evidence for $H_0:\delta=0$ against the evidence for $H_1:\delta \neq 0$ (or $H_1:\delta < 0$, or $H_1:\delta > 0$) given the data $x$. In the case when $BF_{10}=20$, $H_1$ is 20 times more likely after observing the data than $H_0$. The natural question following in such cases is: How large is $\delta$? A Bayes factor cannot answer this question and was not designed to answer such questions, but often this is of most relevance in applied biomedical research. Last, in most applied research, estimation of the effect size $\delta$ is more desirable than a mere rejection or acceptance of a point or composite hypothesis \cite{Kelter2020BayesianPosteriorIndices}, \cite{Kelter2020}, \cite{Kruschke2018}. 

To be fair enough, Bayes factors can be computed alongside posterior estimates, so testing and estimation do not mutually exclude each other. As the Bayes factor is often proposed as a replacement for the p-value, it is even more questionable if practitioners will really combine testing with estimation of effect sizes, especially when scales translating the size of a Bayes factor into evidence (similar to $p<.05$, $p<.01$) are regularly provided by now, see \cite{VanDoorn2019} and \cite{Bergh2019ANOVAJasp}. p-values and the Bayes factor are a useful tools if explicit hypothesis testing is necessary. However, exactly this necessity may be questioned in a wide range of applied biomedical research. 

Therefore, this paper proposes an alternative Bayesian two-sample t-test by formulating the statistical model as a two-component Gaussian mixture with known allocations and using the region of practical equivalence (ROPE). Instead of focussing on rejection or confirmation of hypotheses, the proposed method's focus lies on estimation of the effect size under uncertainty.

\section{Method}
\label{sec2}

\subsection{Modeling the Bayesian t-test as a mixture model with known allocations}\label{sec:bayesianTTestAsMixtureModel}

In this section, the two-sample t-test is modelled as a two-component Gaussian mixture with known allocations.
\begin{quote}
	\textit{``Consider a population made up of $K$ subgroups, mixed at random in proportion to the relative group sizes $\eta_1,...,\eta_K$. Assume interest lies in some random feature $Y$ which is heterogeneous across and homogeneous within the subgroups. Due to heterogeneity, $Y$ has a different probability distribution in each group, usually assumed to arise from the same parametric family $p(y|\theta)$ however, with the parameter $\theta$ differing across the groups. The groups may be labeled through a discrete indicator variable $S$ taking values in the set $\{1,...,K\}$.\newline
	When sampling randomly from such a population, we may record not only $Y$, but also the group indicator $S$. The probability of sampling from the group labeled $S$ is equal to $\eta_S$, whereas conditional on knowing $S$, $Y$ is a random variable following the distribution $p(y|\theta_S)$ with $\theta_S$ being the parameter in group $S$. (...) The marginal density $p(y)$ is obviously given by the following mixture density
	\begin{align*}
		p(y)=\sum_{S=1}^K p(y,S)=\eta_1 p(y|\theta_1)+...+\eta_K p(y|\theta_K)	
	\end{align*}''}
	\cite[p.~1]{Fruhwirth-Schnatter2006}
	\end{quote}
	Clearly, this resembles the situation of the two-sample t-test, in which the allocations $S$ are known. While traditionally mixtures are treated with missing allocations, in the setting of the two-sample t-test these are known, leading to a ``degenerate'' mixture\footnote{The mixture is called degenerate here, because when allocations are known, the likelihood is not mixed in the classical sense.}. While this assumption does not only remove computational difficulties like label switching, it also makes sense from a semantic perspective: the inherent assumption of a researcher is that the population is indeed made up of $K=2$ subgroups, which differ in a random feature $Y$ which is heterogeneous across groups and homogeneous in each group. The group indicator $S$ of course is recorded. When conducting a randomized controlled trial (RCT), the clinician will choose the patients according to a sampling plan, which could be set to achieve equally sized groups, that is, $\eta_1=\eta_2$. Therefore, when sampling the population with the goal of equally sized groups, the researcher samples the objects with equal probability from the population. %In practice, consider a treatment and a control group. The researcher could flip a coin for each patient in the study to assign him or her to the treatment or control group, so that with probability $\eta_1=0.5$ the patient is assigned to the treatment group, and with probability $\eta_2=0.5$ to the control group. Repeating this process leads to the mixture model given above. 
	After the RCT is conducted, the resulting histogram of observed $Y$ values will take the form of the mixture density $p(y)$ above and express bimodality due to the mixture model of the data-generating process.\footnote{If unbalanced groups are the goal, the weights could be adjusted accordingly. As in most cases equally sized groups are considered, $\eta_1=\eta_2=0.5$ is a justified assumption regarding the sampling process in the study or experiment conducted, when balanced groups are used.} After fixing the mixture weights, the family of distributions for the single groups needs to be chosen. The above considerations lead to consider finite mixtures of normal distributions, as these \textit{`occur frequently in many areas of applied statistics such as [...] medicine'} \cite[p.~169]{Fruhwirth-Schnatter2006}. The components $p(y|\theta_i)$ become $f_N(y;\mu_i,\sigma_i^2)$ for $i=1,...,K$ in this case, where $f_N(y;\mu_i,\sigma_i^2)$ is the density of the univariate normal distribution. Parameter estimation in finite mixtures of normal distributions consists of estimation of the component parameters $(\mu_i,\sigma_i^2)$, the allocations $S_i,i=1,...,n$ and the weight distribution $(\eta_1,...,\eta_K)$ based on the available data $y_i,i=1,...,n$. In the case of the two-sample Bayesian t-test, the allocations $S_i$ (where $S_i=0$ if $y_i$ belongs to the first component and $S_i=1$ else) are known for all observations $y_i$, $i=1,...,n$. Also, the weights $\eta_1,\eta_2$ are known. Therefore, inference is concerned with the component parameters $\mu_k,\sigma_k^2$ given the complete data $S,y$.	

\begin{definition}[Bayesian two-sample t-test model]
	Let $S$, $Y$ be random variables with $S$ taking values in the set $\{1,2\}$ and $Y$ in $\mathbb{R}$. If $Y|S=i \sim \mathcal{N}(\mu_i,\sigma_i^2)$ for $i=1,2$, so conditional on $S$ the component densities of $Y$ are Gaussian with unknown parameters $\mu_i$ and $\sigma_i^2$, and if the marginal density is a two-component Gaussian mixture with known allocations
	\begin{align*}
		p(y)=\eta_1 f_N(y;\mu_1,\sigma_1^2)+\eta_2 f_N(y;\mu_2,\sigma_2^2)
	\end{align*}
	where $\eta_2:=\frac{1}{n}\sum_{i=1}^n \mathds{1}_{S_i = 1}(y_i,S_i)$ and $\eta_1 = 1-\eta_2$, the complete data $S,Y$ follow the Bayesian two-sample t-test model.
	
\end{definition}

\subsection{Inference via Gibbs Sampling}

From the above line of thought it is clear that due to the representation via a mixture model with known allocations, no prior is placed directly on the effect size $\delta:=\frac{\mu_1-\mu_2}{s}$ itself, where
\begin{align*}
s:=\sqrt{\frac{(n_1-1)s_1^2+(n_2-1)s_2^2}{n_1+n_2-2}}	
\end{align*}
and $s_1^2$ and $s_2^2$ are the empirical variances of the two groups, see also Cohen \cite{cohen_statistical_1988}. This is the common approach in existing Bayesian t-tests \cite{Gronau}. Instead, in the proposed mixture model, priors are assigned to the parameters of the Gaussian mixture components $\mu_1,\mu_2$ and $\sigma_1^2,\sigma_2^2$. This has several benefits: Incorporation of available prior knowledge is easier achieved with the mixture component parameters than for the effect size, which is an aggregate of these component parameters. Consider a drug where from biochemical properties it can safely be assumed that the mean in the treatment group will become larger, but the variance will increase, too. Incorporating such knowledge on $\mu_i$ and $\sigma_i$ is much easier than incorporating it in the prior of the effect size $\delta$. This situation holds in particular, when group sizes $n_1, n_2$ are not balanced.

These practical gains of translating prior knowledge into prior parameters comes at a cost: In contrast to existing solutions \cite{Gronau}, the model implies that no closed form expression for the posterior of $\delta$ is available. Therefore, sampling methods are used here, to first construct the joint posterior $p(\mu_1,\mu_2,\sigma_1,\sigma_2|S,y)$ and subsequently use a sample
\begin{align*}
	(\mu_1^{(1)},\mu_2^{(1)},\sigma_1^{(1)},\sigma_2^{(1)},...,\mu_1^{(m)},\mu_2^{(m)},\sigma_1^{(m)},\sigma_2^{(m)})
\end{align*}
of size $m$, to produce a sample $(\delta^{(1)},\delta^{(2)},...,\delta^{(m)})$ of $\delta$, where $\delta^{(i)}:=\frac{\mu_1^{(i)}-\mu_2^{(i)}}{s^{(i)}}$ and 
\begin{align*}
	s^{(i)}=\sqrt{\frac{(n_1-1)(s_1^{(i)})^2+(n_2-1)(s_2^{(i)})^2}{n_1+n_2-2}}
\end{align*}
In summary, via Gibbs sampling the posterior of $\delta$ can be approximated reasonably well. In order to apply Gibbs sampling, the conditional distributions need to be derived.

\subsection{Derivation of the full conditionals using the independence prior}\label{subsec:derivationFullConditionalsAndPriors}

To derive the full conditionals, it first has to be decided which priors should be used on the mixture component parameters. There are multiple priors available, the most prominent among them the conditionally conjugate prior and the independence prior \cite{Escobar1995, Fruhwirth-Schnatter2006}. While the conditionally conjugate prior has the advantage of leading to a closed-form posterior $p(\mu,\sigma^2|S,y)$, the main difficulty in the setting of the Bayesian two-sample t-test is that while a priori the component parameters $\theta_k=(\mu_k,\sigma_k^2)$ are pairwise independent across both groups, inside each group the mean $\mu_k$ and variance $\sigma_k^2$ are dependent. This is in contrast to the assumption in the setting of the Bayesian two-sample t-test, and therefore the independence prior is chosen, which is used in \cite{Escobar1995} and \cite{RichardsonGreen1997}. The independence prior assumes the mean $\mu_k$ and the variance $\sigma_k^2$ are a priori independent, that is $
	p(\mu,\sigma^2)=\prod_{k=1}^K p(\mu_k) \prod_{k=1}^K p(\sigma_k^2)	$,
	with $\mu_k \sim \mathcal{N}(b_0,B_0)$ and $\sigma_k^2 \sim IG(c_0,C_0)$, where $IG(\cdot)$ is the inverse Gamma distribution. The normal prior on the means $\mu_k$ seems reasonable as the parameters $b_0$ and $B_0$ can be chosen to keep the influence of the prior only weakly informative.\footnote{Another option would be a $t_n$ prior, but this would also imply another free hyperparameter to be estimated simultaneously, the degrees of freedom $n$.} The inverse Gamma prior is chosen because for a two-component Gaussian mixture to show any signs of bimodality -- in which case one would assume differences between two subgroups in the whole sample -- the variance should not be huge, because otherwise the modes (or the bell-shape) of the two normal-components of the mixture will flatten out more and more, until unimodality is reached. Thus, the inverse Gamma prior connects this model aspect by giving more probability mass to smaller values of $\sigma_k^2$, while extremely large values get much less prior probability mass.\footnote{Another option would be an exponential prior with parameter $\lambda$, but as the exponential distribution is just a special case of the gamma distribution, the more general gamma prior is selected here.} The hyperparameters $c_0$ and $C_0$ then offer control over this kind of shrinkage on $\sigma_k^2$ towards zero. In the simulation study below the prior sensitivity will also be studied briefly.
	 
The independence prior is therefore used and leads to the following full conditionals:
\begin{theorem}\label{theorem:fullConditionals}
	For the Bayesian two-sample t-test model, the full conditional distributions under the independence prior
	\begin{align*}
		p(\mu,\sigma^2)=\prod_{k=1}^K p(\mu_k) \prod_{k=1}^K p(\sigma_k^2)
	\end{align*}
	with $\mu_k \sim \mathcal{N}(b_0,B_0)$ and $\sigma_k^2 \sim IG(c_0,C_0)$ (where $IG(\cdot)$ is the inverse Gamma distribution) are given as:
\begin{align*}
		&p(\mu_1|\mu_2,\sigma_1^2,\sigma_2^2,S,y)=p(\mu_1|\sigma_1^2,S,y)\sim \mathcal{N}(b_1(S),B_1(S))\\
		&p(\mu_2|\mu_1,\sigma_1^2,\sigma_2^2,S,y)=p(\mu_2|\sigma_2^2,S,y)\sim \mathcal{N}(b_2(S),B_2(S))\\
		&p(\sigma_1^2|\mu_1,\mu_2,\sigma_2^2,S,y)=p(\sigma_1^2|\mu_1,S,y)\sim IG(c_1(S),C_1(S))\\
		&p(\sigma_2^2|\mu_1,\mu_2,\sigma_1^2,S,y)=p(\sigma_1^2|\mu_1,S,y)\sim IG(c_2(S),C_2(S))
\end{align*}
with $B_1(S),b_1(S),B_2(S),b_2(S)$ as defined in equations (\ref{eq:Bk}) and (\ref{eq:bk}), and $c_1(S),c_2(S),C_1(S)$ and $C_2(S)$ as defined in equations (\ref{eq:ck}) and (\ref{eq:Ck}) in the Appendix A.2.
\end{theorem}
A proof is given in Appendix A.2, which builds on the derivations in Appendix A.1.
Note that when $\eta_1 \neq \eta_2$, $N_1(S)$ and $N_2(S)$ in the Appendices just need to be changed accordingly. For example, if the first group consists of $30$ observations, and the second group of $70$, setting $N_1(S)=30$ and $N_2(S)=70$ implies $\eta_1=0.3$ and $\eta_2=0.7$, handling the case of unequal group sizes easily.

\subsection{Derivation of the single-block Gibbs sampler}

Based on the full conditionals derived in the last section, this section now derives a single-block Gibbs sampler to obtain the joint posterior distribution 
\begin{align*}
	p(\mu_1,\mu_2,\sigma_1^2,\sigma_2^2|S,y)
\end{align*}
given the complete data $(S,y)$. The resulting Gibbs sampler is given as follows:\newline
\begin{corollary}[Single-block Gibbs sampler for the Bayesian two-sample t-test]\label{corollary:gibbsSampler} The joint posterior distribution 
	\begin{align*}
	p(\mu_1,\mu_2,\sigma_1^2,\sigma_2^2|S,y)
\end{align*}
in the Bayesian two-sample t-test model can be simulated under the independence prior as:\newline
%\textit{Algorithm 1: Single-Block Gibbs sampler for a univariate Gaussian Mixture of two components with known allocations}\newline
\textit{Conditional on the classification $S=(S_1,...,S_N)$:
\begin{enumerate}
	\item{Sample $\sigma_k^2$ in each group $k$, $k=1,2$ from an inverse Gamma distribution
		$\mathcal{G}^{-1}(c_k(S),C_k(S))$}
	\item{Sample $\mu_k$ in each group $k$, $k=1,2$, from a normal distribution $\mathcal{N}(b_k(S),B_k(S))$}
\end{enumerate}
where $B_k(S), b_k(S)$ and $c_k(S), C_k(S)$ are given by equations (\ref{eq:Bk}), (\ref{eq:bk}), (\ref{eq:ck}) and (\ref{eq:Ck}) in the Appendix.}
\end{corollary}
A proof is given in Appendix A.3.

\subsection{The shift from hypothesis testing to estimation under uncertainty and the ROPE}\label{sec:shiftToNewStatisticsAndROPE} 
%p-values are not useful when it comes to statements about the experiment or study at hand, while Bayes factors -- while useful if testing is necessary -- seem to be mainly a Bayesian replacement of frequentist test levels or p-values proceeding with the same masking of the full information in a single numerical value. Clinicians are often left without any good advice what to do in this situation. 
As already mentioned, instead of explicit hypothesis testing via p-values and Bayes factors, we follow the proposed shift from hypothesis testing to estimation under uncertainty. Cumming \cite{Cumming2014} proposed such a shift originally from frequentist hypothesis testing to estimation, called the \textit{'New Statistics'}, a process observable in a broad range of scientific fields, see Wasserstein \& Lazar \cite{Wasserstein2019}. Note that one of the six principles for properly interpreting $p$-values in the 2016 ASA-statement stressed that a p-value \textit{``does not measure the size of an effect or the importance of a result.''} \cite[p.~132]{wasserstein2016}. Cumming \cite{Cumming2014} therefore included in his proposal a focus on \textit{``estimation based on effect sizes''}  \cite[p.~7]{Cumming2014}. To promote this shift, Kruschke \& Liddell \cite{Kruschke2018} offered two conceptual distinctions, which are replicated in table 1 below.

While Cumming \cite{Cumming2014} originally proposed a shift from frequentist hypothesis testing to frequentist estimation under uncertainty, we propose that this shift can be achieved easier by Bayesian methods. The main reasons are that confidence intervals as quantities for estimation are still \textit{``highly sensitive to the stopping and testing intentions.''} \cite[p.~184]{Kruschke2018}, while Bayesian posterior distributions are not, see also Berger \&  Wolpert \cite[Chapter 4]{Berger1988a}.

\begin{table}[h!]
  \caption{Two conceptual distinctions in the practice of data analysis, replicated from Kruschke \& Liddell \cite{Kruschke2018}}
  \label{tab:categorizationKruschke2018}
  \centering
  \begin{tabular}{p{2cm}|p{2.5cm}|p{2.5cm}}
    \hline
	& Frequentist & Bayesian \\ 
	\hline
	Hypothesis test & p-value (null hypothesis significance test) & Bayes factor\\
	\hline
	Estimation with uncertainty & MLE with CI (The ``New Statistics'') & Posterior Distribution with highest posterior density interval (HPD) \\
	\hline
\end{tabular}
\end{table}

%\footnote{The intuitively reasonable irrelevance of stopping rules follows only if the sufficiency principle (SP) and the weak conditionality principle (WCP) are accepted, so that the (relative) likelihood principle (RLP) follows. It is argued that most practitioners do accept the SP and WCP, so that Bayesian methods are coherent from a foundational perspective, while NHST is not, see \cite[Chapter 4]{Berger1988a}.}. 

%In summary, maximum likelihood estimates (MLE) combined with confidence intervals (CI) still suffer from multiple problems inherent in frequentist NHST -- which is because of the duality between tests and confidence sets, while Bayesian posterior distributions with highest density intervals do not.

\subsection{The proposal of a region of practical equivalence}
To facilitate the shift to an estimation-oriented perspective, Kruschke \& Liddell \cite{Kruschke2018} advertised the \textit{region of practical equivalence (ROPE)}. As they note: \textit{`ROPE's go by different names in the literature, including ``interval of clinical equivalence'', ``range of equivalence'', ``equivalence interval'', ``indifference zone''}, \textit{``smallest effect size of interest,''} \textit{ and ``good-enough belt'' ...'} \cite[p.~185]{Kruschke2018}, where these terms come from a wide spectrum of scientific domains, see Carlin \& Louis \cite{Carlin2009}, Freedman, Lowe and Macaskill \cite{Freedman1983}, Hobbs \& Carlin \cite{Hobbs2007}, Lakens \cite{Lakens2014} and Schuirmann \cite{Schuirmann1987}. The uniting idea is to establish a region of practical equivalence around the null value of the hypothesis, which expresses \textit{``the range of parameter values that are equivalent to the null value for current practical purposes.''} \cite[p.~185]{Kruschke2018}. With a caution not to slip back into dichotomic black-and-white thinking, the following decision rule was proposed by Kruschke \& Liddell \cite{Kruschke2018}: Reject the null value, if the 95\% highest posterior density interval (HPD) falls entirely outside the ROPE. Accept the null value, if the 95\% HPD falls entirely inside the ROPE. In the first case, with more than 95\% probability the parameter value is not inside the ROPE, and therefore not practically equivalent to the null value. A rejection of the null value then seems legitimate. In the second case, the parameter value is inside the ROPE with at least 95\% posterior probability, and therefore practically equivalent to the null value. It seems legitimate to accept the null value. Of course, it would also be possible to accept the null value iff the whole posterior is located inside the ROPE, leading to an even stricter decision rule.

%Note that in this case accepting the null value is equal to accepting \textit{any} value inside the ROPE, because by definition of the ROPE, these values are practically equivalent. If for example in the setting of the Bayesian t-test, a ROPE of $(-0.2,0.2)$ is chosen around the null value $\delta=0$ (where $\delta$ is the effect size, defined as $\delta:=\frac{\mu_1-\mu_2}{s}$ where $s=\sqrt{\frac{(n_1-1)s_1^2+(n_2-1)s_2^2}{n_1+n_2-2}}$, which becomes $\frac{\mu_2-\mu_1}{\sqrt{(\sigma_1^2+\sigma_2^2)/2}}$ when $n_1=n_2$) and the 95\% HPD of the posterior of $\delta$ lies inside this ROPE, accepting the null value $\delta=0$ means to accept that $\delta$ has \textit{any} value inside the ROPE, which are interpreted as equivalent for practical purposes. Therefore, $\delta=0$ cannot be distinguished from e.g. $\delta=0.05$ or $\delta=-0.1$. 

While the idea of a ROPE is intriguing, a limitation should be noted which is that it can only apply in situations where scientific standards of practically equivalent parameter values exist and are widely accepted by researchers. Luckily, this is the case for effect sizes, which have a long tradition of being categorised in biomedical and psychological research, see Cohen \cite{cohen_statistical_1988}. 

\begin{definition}[ROPE]
	The region of practical equivalence (ROPE) $R$ for (or around) a hypothesis $H\subset \Theta$ is a subset of the parameter space $\Theta$ with $H\subset R$.
\end{definition}
A statistical hypothesis $H$ is now described via a region of practical equivalence $R$, e.g. $H:\delta=\delta_0$ can be described as $R:=[\delta_0-\varepsilon,\delta_0+\varepsilon]$ for $\varepsilon >0$. By definition, any set $R\subset \Theta$ with $H\subset R$ is allowed to describe $H$, and should be selected depending on how precise the measuring process of the experiment or study is assumed to be. Next, we define two options for the ROPE:
\begin{definition}[Correctness]
	Let $R\subset \Theta$ a ROPE around a hypothesis $H\subset \Theta$, that is $H\subset R$, where $H$ makes a statement about the unknown model parameter $\theta$. If the true parameter value $\theta_0$ lies in $R$, that is $\theta_0\in R$, then $R$ is called correct, otherwise incorrect.
\end{definition}
A correct ROPE therefore contains the true parameter value $\theta_0$, while an incorrect one does not. 

\subsection{The proposal for a shift towards estimation under uncertainty}
The two major drawbacks of the proposal of Kruschke \cite{Kruschke2018a} and Kruschke \& Liddell \cite{Kruschke2018} are that the ROPE still facilitates hypothesis testing, enforcing a binary decision of rejection or acceptance, while it is also unclear what to do when the 95\%-HPD lies partly inside and partly outside the ROPE. Therefore, a different proposal is made in this paper, which is estimation of the \textit{mean probable effect size (MPE)} instead of hypothesis testing. This procedure will be used in the proposed t-test afterwards. First, the acceptance or rejection of a hypothesis $H$ can be formalized as follows:
\begin{definition}[$\alpha$-accepted / $\alpha$-rejected]
	Let $\theta$ the unknown parameter (or vector of unknown parameters) in an experiment $E:=\{X,\theta, \{f_{\theta}\}\}$, where the random variable $X$ taking values in $\mathbb{R}$ and having density $f_{\theta}$ for some $\theta \subset \Theta$, is observed. Let $f(\theta|x)$ the posterior distribution of $\theta$ (under any prior $\pi(\theta)$), and let $C_{\alpha}$ the corresponding $\alpha$\% highest density interval of $f(\theta|x)$. Let $R\subset \Theta$ a ROPE around the hypothesis $H$ of interest, which makes a statement about $\theta$. Then, if $C_{\alpha} \subset R$, the hypothesis $H$ is called $\alpha$-accepted, else $\alpha$-rejected. If $\alpha=1$, then $H$ is simply called accepted, else rejected.
\end{definition}
Thus, if $C_{\alpha}$ lies completely inside the ROPE $R$ and if $\alpha=1$, the entire posterior probability mass indicates that $\theta$ is practically equivalent to the values described by the ROPE $R$. Thus, $H$ can be accepted. If $\alpha <1$, the strength of this statement becomes less with decreasing $\alpha$ of course. For example, if $H$ is $0.75$-accepted for a given ROPE $R$, 25\% of the posterior indicate that $\theta$ may take values different than the ones included in the ROPE $R$. It is clear that little is gained if the value of $\alpha$ is small or close to zero when speaking of $\alpha$-acceptance. Therefore, instead of forcing an acceptance or rejection (which only makes sense for substantial values of $\alpha$), a perspective focussing on continuous estimation is preferred: 
\begin{definition}[Posterior mass percentage]
	Let $f(\theta|x)$ a posterior for $\theta$ and $\delta_{MPE}:=\mathbb{E}[\theta|x]$ the mean posterior effect size (where the expectation $\mathbb{E}$ is taken with respect to the posterior). Let $R_1,...,R_m$ be a partition of the support of the posterior $f(\theta|x)$ into different ROPEs corresponding to different hypotheses $H_1,...,H_m$, which make statements about the unknown parameter (vector) $\theta$. Without loss of generality, let $R_j$ the ROPE for which $\delta_{MPE} \subset R_j$, $j \subset \{1,...,m\}$. The posterior mass percentage $PMP_{R_j}(\delta_{MPE})$ of $\delta_{MPE}$ is given as
	\begin{align*}
		PMP_{R_j}(\delta_{MPE}):=\int_{R_j} f(\theta|x)d\theta
	\end{align*}
	that is, the percentage of the posterior distribution's probability mass inside the ROPE $R_j$ around $\delta_{MPE}$. 
\end{definition}
For simplicity of notation, the subscript $R_j$ is omitted whenever it is clear which ROPEs $R_j$ are used for partitioning the support of $f(\theta|x)$. Now, in contrast to strict $\alpha$-acceptance or $\alpha$-rejection rules based on the ROPE $R_j$, we propose to use $\delta_{MPE}$ and $PMP(\delta_{MPE})$ together to estimate the effect size $\delta$ under under uncertainty, and to quantify this uncertainty via $PMP(\delta_{MPE})$. If $\delta_{MPE}$ is non-zero, the t-test found a difference between both groups. The size of this difference is quantified by $\delta_{MPE}$ itself. The uncertainty in this statement is quantified by $PMP(\delta_{MPE})$. For the developed two-sample t-test, we propose the following procedure:
	\begin{enumerate}
		\item{For a fixed credible level $\alpha$, the \textit{effect size range (ESR)} should be reported. That is, which effect sizes $\delta$ are assigned positive probability mass by the $\alpha$\% HPD interval, $0\leq \alpha \leq 1$. The ESR is a first estimate of credible effect sizes a posteriori.}
		\item{The support of the posterior distribution $f(\delta|S,Y)$ in the Bayesian t-test model is partitioned into the standardized ROPEs of the effect size $\delta$ of \cite{cohen_statistical_1988}, leading to a partition $\mathcal{P}$ of the support as given in the definition of $PMP(\delta_{MPE})$.}
		\item{The \textit{mean posterior effect size} $\delta_{MPE}$ is calculated as an estimate of the true effect size $\delta$. The surrounding ROPE $R_j$ with $\delta_{MPE} \subset R_j$ of the partition $\mathcal{P}$ is selected, and the exact percentage inside $R_j$ is reported as the \textit{posterior mass percentage} $PMP(\delta_{MPE})$.}
	\end{enumerate}
	 %Note that while the posterior density is dominated by the Lebesgue measure in the case of the two-sample t-test, the posterior probability mass of a single point is zero, but the posterior density value is not, so the maximum $\delta_{MPE}$ can be calculated.
	 The above procedure leads to an estimation of the effect size $\delta$ under uncertainty instead of a hypothesis testing perspective. Additionally to the posterior mean $\delta_{MPE}$, the posterior mass percentage $PMP(\delta_{MPE})$ gives a \textit{continuous} measure of the trustworthiness of the estimate ranging from $0$\% to $100$\% (actually from zero to one, but for better interpretability how much of the posteriors mass is allocated in the ROPE $R_j$ the values zero to 100 percent will be used in what follows). $\delta_{MPE}$ estimates with $PMP(\delta_{MPE})>0.5$ (or 50\%) could be interpreted as decisive, but do not need to. $PMP(\delta_{MPE})$ can be treated as a continuous measure of support for the effect size estimated by $\delta_{MPE}$. There are multiple advantages of utilising a ROPE and combining it with $\delta_{MPE}$ and $PMP(\delta_{MPE})$, the most important of which may be:
	 \begin{theorem}\label{theorem:consistencyDeltaMPE}
	 	Let $R_j\subset \Theta$ a ROPE around $\delta_{MPE}$, that is $\delta_{MPE} \subset R_j$. If $R_j$ is correct, then $PMP(\delta_{MPE})\rightarrow 1$ for $n\rightarrow \infty$ almost surely, and if $R_j$ is incorrect, then $PMP(\delta_{MPE})\rightarrow 0$ for $n\rightarrow \infty$ almost surely, except possibly on a set of $\pi$-measure zero for any prior $\pi$ on $\theta$.
	 \end{theorem}
A proof is given in Appendix A.4. Using $\delta_{MPE}$ together with $PMP(\delta_{MPE})$ therefore will eventually lead to the correct estimation of $\delta$ in the sense that when a correct ROPE is chosen, the posterior mass percentage will converge to one, and if an incorrect ROPE is chosen, the posterior mass percentage will converge to zero. Thus, the procedure indicates whether $\delta$ is practically equivalent to the values given by the ROPE or not. If necessary, explicit hypothesis testing can be performed via $\alpha$-rejection. The advantages compared to p-values and Bayes factors which favour an explicit hypothesis testing perspective are:
	\begin{enumerate}
		\item{As \cite[p.~338]{Greenland2016} stress with regard to the dichotomy induced by hypothesis testing, \textit{`estimation of the size of effects and the uncertainty surrounding our estimates will be far more important for scientific inference and sound judgment than any such classification.'}}
		\item{In contrast to the Bayes factor (BF) the ROPE and $\delta_{MPE}$ have important advantages: they do not encourage the same automatic calculation routines as Bayes factors. For example, \cite{Gigerenzer2015} warned explicitly against Bayes factors becoming the new $p$-values due to the same automatic calculation routines, and the approach via the ROPE fosters estimation and judging the evidence based on the \textit{continuous} support for $\delta_{MPE}$ provided by $PMP(\delta_{MPE})$, instead of using thresholds.}
		\item{The ROPE is supported by the following argument, which questions the use of testing point hypothesis like $H_0:\delta=0$ (no matter if rejecting or confirming is the goal): In practice, measuring is always done with finite precision (like blood pressure, or the heart rate), and therefore the goal rarely is to show (or reject) that the effect size $\delta$ is exactly \textit{equal} to zero, but much more that $\delta$ is \textit{negligibly small} to deny the existence of any existing, (clinically) relevant effect. Therefore, invariances like $\delta=0$ can be interpreted as not existing, at least not exactly, and the search for approximate invariances, as described by a ROPE $R=(-.2,.2)$ around $\delta=0$ is intellectually (more) compelling. A clinician will be satisfied by the statement that the true effect size is not exactly zero, but with 95\% probability negligibly small.}	
	\end{enumerate}

%	\begin{quote}
%		\textit{``... it is worth considering the argument that invariances do not exist, at least not exactly. Cohen (1994), for example, started with the proposition that all variables affect all others to some, possibly small, extent. Fortunately, there is no real contradiction between adhering to Cohen's view that invariances cannot hold exactly for relatively trivial reasons that are outside the domain of study. When they hold only approximately, they often provide a more parsimonious description of data than do the alternatives and can serve as guidance for theory development. Hence, wether one believes that invariances may hold exactly or only approximately, the search for them is intellectually compelling.''}\newline
%		\cite[p.~226]{Rouder2009}
%	\end{quote}

\subsection{Illustrative example}\label{sec:illustrativeExample}	
\begin{figure}[h!]
    \centering
    \includegraphics[width=0.5\textwidth]{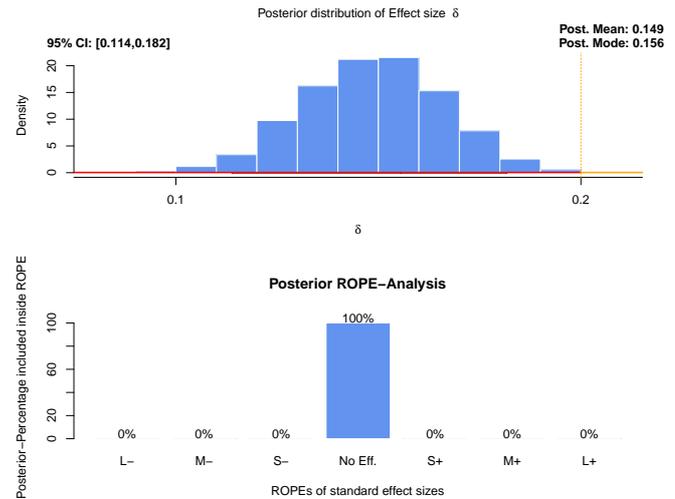}
    \caption{Posterior distribution of the effect size $\delta$ and analysis for the kitchen roll RCT of \cite{Wagenmakers2015} via $\delta_{MPE}$ and $PMP(\delta_{MPE})$}
    \label{fig:ROPEAnalysis}
\end{figure}

To clarify the above line of thought, the following example combines the developed Gibbs sampler for the Bayesian t-test with the ROPE, $\delta_{MPE}$ and $PMP(\delta_{MPE})$. It uses data from Wagenmakers et al. \cite{Wagenmakers2015}, who conducted a randomized controlled trial in which participants had to fill out a personality questionnaire while rolling a kitchen roll clockwise or counter-clockwise. The mean score of both groups was compared afterwards. A traditional two-sided two-sample Welch's t-test indicates that there is no significant difference between both groups, yielding a p-value of $0.4542$.\footnote{The data and code including a full replication script for all simulations and figures presented in the paper is available at \url{https://osf.io/cvwr5/}} What is missing is the effect size, which is of much more interest. Note that computing the effect size from the raw study data does not quantify the uncertainty in the data, which is undesirable. From the p-value, a clinician can only judge that the results are unlikely to be observed under the null hypothesis. However, if the effect is clinically relevant or negligible remains unknown (or at best only based on the raw sample effect size). In this case, as the p-value is quite large, neither can the null hypothesis of no effect be rejected, nor can be said with certainty that there is indeed no effect in the sense of confirming the null hypothesis, leaving the clinician without any trace to proceed with but to collect more data. Figure \ref{fig:ROPEAnalysis} in contrast shows an analysis of the posterior of $\delta$ produced by the Gibbs sampler for the Bayesian t-test. In it, the ROPE, $\delta_{MPE}$ and $PMP(\delta_{MPE})$ are used.
The posterior distribution of $\delta$ is given in the upper plot and shows that the posterior mean is $0.149$ and the posterior mode $0.156$, that is, the mean posterior effect size given the data is $0.149$, no effect discernible from zero. The 95\% highest density interval ranges from $0.114$ to $0.182$, showing that with 95\% probability, there is no effect discernible from $\delta=0$, given the data. Even when taking the 100\% highest posterior density interval (HPD), this situation does not change as indicated by the upper plot. The coloured horizontal lines and vertical dotted lines represent the boundaries of the different ROPEs according to Cohen \cite{cohen_statistical_1988}. The lower plot shows the results of partitioning the posterior mass of $\delta$ into the ROPEs for different effect sizes, which are standardized as small, if $\delta \in [0.2,0.5)$, medium, if $\delta \in [0.5,0.8)$ and large, if $\delta \in [0.8,\infty)$ \cite{cohen_statistical_1988}. $100$\% of this posterior probability mass lies inside the \textit{ROPE} $(-0.2,0.2)$ of no effect. So $\delta_{MPE}=0.149$ indicates that no effect discernible from zero is apparent, and the posterior mass percentage $PMP(\delta_{MPE})=1$ (or 100\%) shows that the estimate $\delta_{MPE}$ is trustworthy, as the entire posterior probability mass is located inside the ROPE $(-0.2,0.2)$ of no effect (also indicated by the upper plot). Based on this analysis, one can conclude that given the data, it is highly probable, that there exists no effect. The method provides more insight than the information a p-value is giving: In the example, the p-value cannot reject the null hypothesis $H_0:\delta = 0$ and neither can the null hypothesis be confirmed. Even if the p-value would have been significant, this means only that the result will unlikely have happened by chance under the null hypothesis. The non-significant p-value of $0.4542$ in this case allows not to accept the null hypothesis of no effect. The proposed procedure in contrast does. Note also that a Bayes factor would have to be combined with estimation to yield the same information, and a Bayes factor alone of course would not have provided this information. In the example, the Bayes factor $BF_{10}$ of $H_1:\delta \neq 0$ against $H_0:\delta=0$ is $BF_{10}=5.015$ when using the recommended wide Cauchy $C(0,1)$ prior of Rouder et al. \cite{Rouder2009}, which indicates only moderate evidence for the null hypothesis $H_0:\delta=0$ according to Van Doorn et al. \cite{VanDoorn2019}. This is in sharp contrast to the $PMP(\delta_{MPE})$ value of $100\%$, which strongly suggests that the null hypothesis $H_0:\delta  =0$ is confirmed. The posterior in figure \ref{fig:ROPEAnalysis} is obtained by the Gibbs sampler given in Corollary 1.

\section{Simulation study}\label{sec:simulationStudy}

Primary interest now lies in the ability to correctly estimate different sizes of effects via the combination of the derived t-test, $\delta_{MPE}$ and $PMP(\delta_{MPE})$. The effect size ROPEs are oriented at the standard effect sizes of Cohen \cite{cohen_statistical_1988}, where an effect is categorized as \textit{small}, if $\delta \in [0.2,0.5)$ or $\delta \in (-0.5,-0.2]$, \textit{medium}, if $\delta \in [0.5,0.8]$ or $\delta \in (-0.8,-0.5]$ and \textit{large}, if $\delta \geq 0.8$ or $\delta \leq -0.8$. Secondary interest lies in analysing if the Gibbs sampler achieves better performance regarding the type I and II error compared with Welch's t-test, the standard NHST solution.\footnote{We do not compare the type I error rate with Bayes factors, as these error types are formally not defined for the Bayes factor.} The plan of the study is as follows: If there is indeed an effect, the Gibbs sampler should lead to a posterior distribution of $\delta$ which lies outside the ROPE $(-0.2,0.2)$, which is equivalent to the rejection of the null hypothesis $H_0:\delta=0$. The precise estimation of the size of an effect is a second task, one more demanding than the sole rejection of $H_0:\delta=0$. If the sampler correctly rejects the null hypothesis, because the posteriors concentrate in the set $(-\infty,-0.2]\cup [0.2,\infty)$, this indicates that it makes no type II error and subsequently achieves a power of nearly 100\%. Of course, this will depend on the sample sizes in both groups. If additionally, the 95\%-credible intervals of the posteriors concentrate in the set $\{(-0.5,-0.2]\cup [0.2,0.5)\}$, then the Gibbs sampler is also consistent for small effect sizes, again depending on the sample size. The same rationale applies for medium effect sizes and the ROPE $\{(-0.8,0.5]\cup [0.5,0.8)\}$ and for large effect sizes and the ROPE $\{(-\infty,0.8]\cup [0.8,\infty)\}$. Therefore, three two-component Gaussian mixtures have been fixed in advance, each representing one of the three effect sizes. For the small effect, the first component is $\mathcal{N}(2.89,1.84)$ and the second component  $\mathcal{N}(3.5,1.56)$, resulting in a effect size of $\delta=(2.89-3.5)/\sqrt{((1.56^2+1.84^2)/2)}=-0.35$. For a medium effect, the first and second group are simulated as $\mathcal{N}(254.08,2.36)$ and $\mathcal{N}(255.84,3.04)$, yielding a true effect size of
	\begin{align*}
			\delta=\frac{(255.84-254.08)}{\sqrt{((3.04^2+2.36^2)/2)}}=0.6467\approx 0.65
	\end{align*}
For the large effect, the first and second group are simulated as $\mathcal{N}(15.01,3.4^2)$ and $\mathcal{N}(19.91,5.8^2)$, yielding a true effect size of
		\begin{align*}
			\delta=\frac{(19.91-15.01)}{\sqrt{((5.8^2+3.4^2)/2)}}=1.03
		\end{align*}
In each of the three effect size scenarios, 100 datasets of the corresponding two-component mixture were simulated for different sample sizes and the Gibbs sampler was run for each of the 100 datasets for 10000 iterations, using a burnin of 5000. $\delta_{MPE}$, the ESR and the ROPE criterion together with $\alpha$-acceptance are applied, that is, the hypothesis $H$ stating a small, medium or large effect size is $\alpha$-accepted if the 95\%-HPD lies completely inside the corresponding ROPE $\{(-0.5,-0.2]\cup [0.2,0.5)\}$, $\{(-0.8,-0.5]\cup[0.5,0.8)\}$ or $\{(-\infty,-0.8]\cup[0.8,\infty)\}$. The recommended wide prior was used for all simulations, which is detailed later in the prior sensitivity analysis.		
		In total, the Gibbs sampler should stabilize around the true effect size $\delta$.\footnote{Note, that here balanced groups are treated, but unbalanced groups could also easily be treated by setting $N_1(S)$ and $N_2(S)$ accordingly, as described above.	}
		
\subsection{Results}
\begin{figure*}[h!]
    \centering
    \label{fig:effSizes}
    \includegraphics[width=1\textwidth]{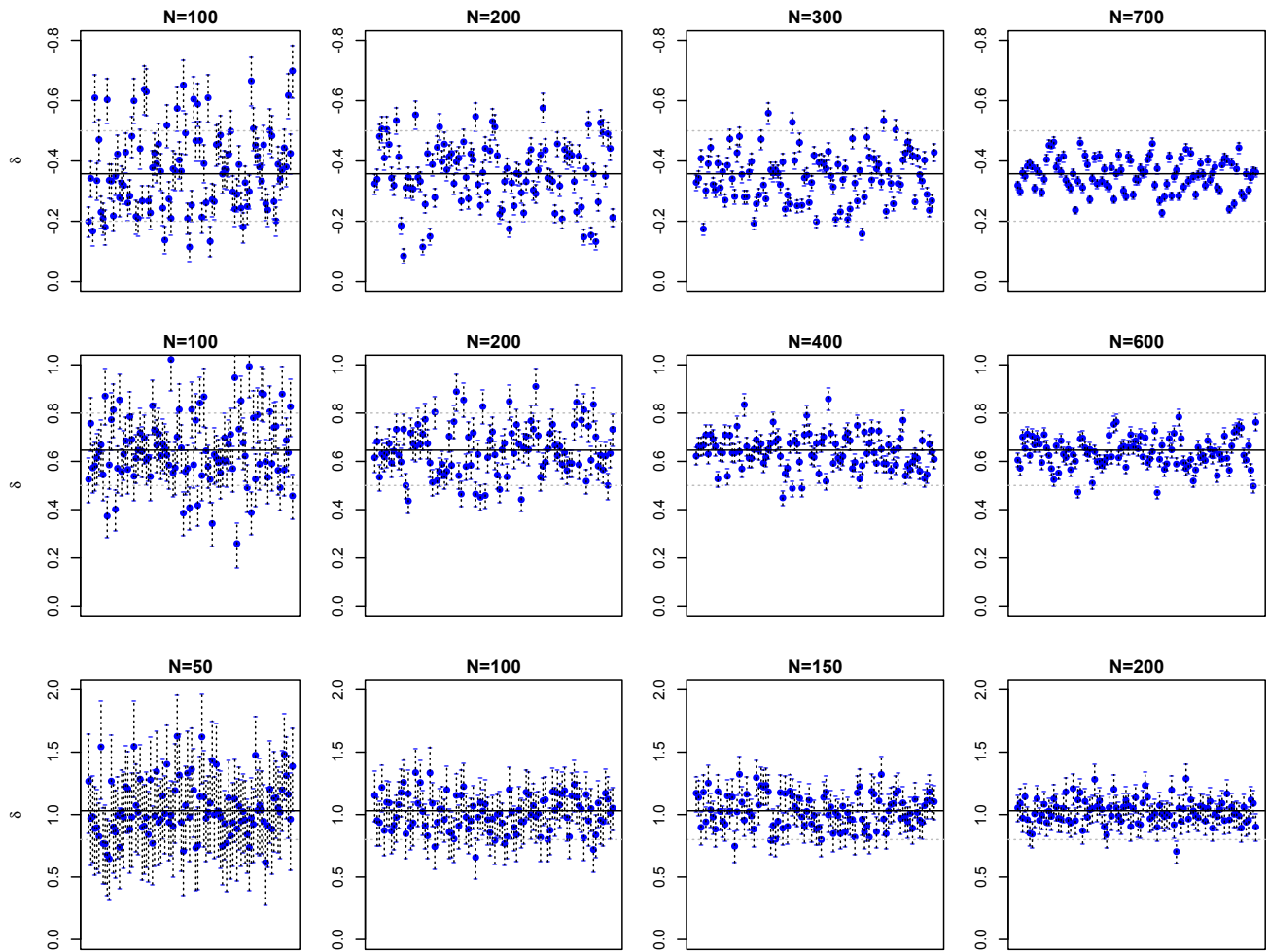}
    \caption{Posterior Means $\delta_{MPE}$ and 95\% credible-intervals for $\delta_{MPE}$ for 100 datasets consisting of sample sizes $2n$, with $n$ observations in each group; dotted lines represent the ROPE boundaries; \textit{upper row}: Small effect size; \textit{middle row}: Medium effect size; \textit{lower row}: Large effect size}
\end{figure*}
	The upper row of Fig. 2 shows the results for small effect sizes. The two left plots show the results for $n=100$ and $n=200$ observations in each group. It is clear that the 95\%-HPDs in both cases fluctuate strongly, indicating that anything from no effect to a medium effect is possible. The two right plots of the upper row show the results when increasing to $n=300$ and $n=700$ observations per group. The 95\%-HPDs get narrower and stabilize inside the ROPE. While for $n=300$ there are still some outliers, for $n=700$ all HPDs have concentrated inside the ROPE of a small effect -- that is $PMP(\delta_{MPE})=1$ (100\%) for all iterations -- and the estimates $\delta_{MPE}$ (blue points) have already converged closely to the true effect size indicated by the solid black line. The necessary sample size for this precision is not small, but a small effect requires a large sample size to be detected.
		
	The middle row of Fig. 2 shows the results for medium effect sizes. The two left plots in it show the result for $n=100$ and $n=200$ observations in each group for a medium effect size. Increasing the sample size to $n=400$ and $n=600$ leads to the results shown in the two right plots. These figures show that even for sample sizes of $n=100$ in both groups, no 95\% HPD lies completely inside the ROPE $(-0.2,0.2)$ around $\delta_0=0$ of no effect, indicating that while the size of the effect may still not be estimated accurately, a null hypothesis of no effect $\delta_0=0$ could always be rejected when using sample sizes of at least $n=100$ in each group and the underlying effect has medium size. When it comes to precisely estimating the size of the effect, larger sample sizes similar to those needed to detect small effect sizes are necessary, as shown by the right plots of the second row.
		
	The lower row of Fig. 2 shows the results for large effect sizes. About $n=50$ observations in each group suffice to produce $\delta_{MPE}$ and $PMP(\delta_{MPE})$ which estimate small to large effects, and thereby reject a null hypothesis of no effect, while about $n=150$ to $n=200$ seem reasonable to precisely estimate a large effect size. When using $\delta_{MPE}$ (blue points) as an estimator for $\delta$, sample sizes of $n=200$ produce an estimate close to the true effect size, which in this case was $\delta_0=1.030723$.

\subsection{Controlling the type I error rate}\label{subsec:controllingTypeIErrors}
In frequentist NHST, the Neyman-Pearson theory aims at controlling the type I error rate $\alpha$, which is the probability to reject the null hypothesis $H_0$ falsely, when indeed it is correct. In the setting of the two-sample Bayesian t-test this equals the rejection of $H_0:\delta=0$ although the true effect size is $\delta_0=0$. Following Cohen \cite{cohen_statistical_1988}, an effect is considered small if the effect size is at least $|\delta|\geq 0.2$, so effect sizes in the interval $(-0.2,0.2)$ can be considered as noise, or \textit{practically equivalent to zero}. Therefore, a ROPE of $(-0.2,0.2)$ is set around the null value $\delta_0=0$ to compare the type I error rate of the proposed method against the standard frequentist NHST solution, Welch's t-test. Again 100 datasets of different sample sizes are simulated where the true effect size $\delta_0$ is set to zero. The Gibbs sampler should produce a posterior distribution of $\delta$ which concentrates inside the ROPE, so that the null hypothesis $H:\delta_0=0$ is $\alpha$-accepted for $\alpha=0.95$, see definition 4. If the 95\%-HPD interval lies (entirely) outside the ROPE, this equals $\alpha$-rejection for $\alpha=0.95$, or in frequentist terms the rejection of the null hypothesis $H_0:\delta_0=0$ of no effect and therefore the commitment of a type I error. The following two definitions formalize the type I and II error building on the concept of $\alpha$-rejection:
\begin{definition}[$\alpha$ type I error]
	An $\alpha$ type I error happens if the true parameter value $\delta_0 \in H$, with $H\subset R$ for a ROPE $R\subset \Theta$, but $H$ is $\alpha$-rejected for $\alpha$. 
\end{definition}
\begin{definition}[$\alpha$ type II error]
	An $\alpha$ type II error happens if the true parameter value $\delta_0 \notin H$ and $\delta_0 \notin R$, with $H\subset R$ for a ROPE $R\subset \Theta$, but $H$ is $\alpha$-accepted for $\alpha$. 
\end{definition}
\begin{figure}[h!]
    \centering
    \includegraphics[width=0.5\textwidth]{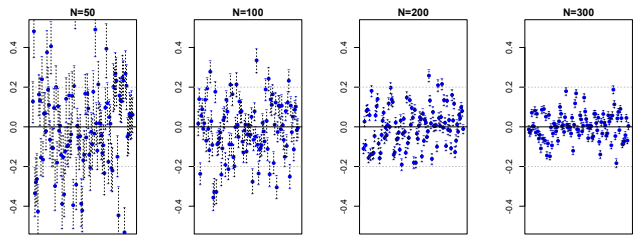}
    \label{fig:noEffect}
    \caption{Posterior means $\delta_{MPE}$ and 95\% credible-intervals for $\delta$ for 100 datasets consisting of different sample sizes $2n$, with $n$ observations from a $\mathcal{N}(148.3,1.34)$ distribution and $n$ observations from a $\mathcal{N}(148.3,2.04)$ distribution; dotted lines represent the ROPE $(-0.2,0.2)$ of no effect size around $\delta=0$; posterior distributions are based on 10000 iterations of the Gibbs sampler with a burnin of 5000 iterations}
\end{figure}
The left plot in figure 3 shows the results of 100 datasets of size $n=50$ in each group. The first group was simulated as $\mathcal{N}(148.3,1.34)$, and the second group as $\mathcal{N}(148.3,2.03)$. The true effect size is
\begin{align*}
	\delta_0=\frac{\mu_2-\mu_1}{\sqrt{(\sigma_1^2+\sigma_2^2)/2}}=0
\end{align*}
The blue points represent $\delta_{MPE}$ and the blue dotted lines the 95\%-HPDs of the posterior of $\delta$. While the estimates fluctuate strongly for $n=50$, increasing sample size in each group successively to $n=200$ as shown by the progression of the plots from left to right shows that false-positive results -- $\alpha$ type I errors with $\alpha=.95$ -- get completely eliminated for sufficiently large sample size. The right plot with sample size $n=300$ shows that no $\alpha$ type I error with $\alpha=.95$ occurs anymore. Also, $\delta_{MPE}$ stabilizes around the true value $\delta_0=0$ of no effect, indicating its convergence to the true effect size $\delta$.

The simulations show that the $\alpha$ type I error rate converges to zero when the sample size is increased. The number of credible intervals which lie partly inside and partly outside the ROPE decreases to zero. In contrast, p-values are uniformly distributed under the null hypothesis, so that no matter what size the samples in both groups are, in the long-run one will still obtain $\alpha$\% (most often 5\%) type I errors. Conducting Welch's t-tests will thus inevitably lead to a type I error rate of 5\%, if the test level is set to $\alpha=.05$. If the sample size is at least $n=200$ in each group, the proposed Bayesian t-test together with $\delta_{MPE}$ and $PMP(\delta_{MPE})$ performs better with respect to control the $\alpha$ type I error rate.

From a theoretical perspective, it is of course of interest for which values of $\alpha$ this fact does hold, and indeed, using the two generalized types of type I and II errors, it can also be shown that the number of type I (type II) errors converges to zero for any $\alpha \neq 0$, when a correct (incorrect) ROPE is chosen:
\begin{theorem}\label{theorem:errors}
	For the Bayesian two-sample t-test model, the probability of making an $\alpha$ type I error for any $\alpha \neq 0$ converges to zero for any correct ROPE $R$ around the hypothesis $H$ which makes a statement about the unknown parameter $\delta$. Also, the probability of making a $\alpha$ type II error for any $\alpha \neq 0$ converges to zero for any incorrect ROPE $R$.
\end{theorem}
A proof is given in Appendix A.5.
 The implications of Theorem 3 are that if a correct ROPE $R$ is chosen, then eventually the probability of making a $\alpha$ type I error will become zero. Thus, when the ROPE $R$ includes the true parameter $\delta_0$, eventually the hypothesis $H$ will be $\alpha$-accepted for $\alpha=1$, that is, accepted. If on the other hand an incorrect ROPE is selected, which does not include the true parameter $\delta_0$, then eventually the probability of making a $\alpha$ type II error -- that is, accepting $H$ although $\delta_0 \notin H$ -- will become zero.

\subsection{Prior sensitivity analysis}
Section \ref{subsec:derivationFullConditionalsAndPriors} detailed the independence prior used in the model, and of specific interest is of course the influence of this prior on the results produced by the procedure. Therefore, three different hyperparameter settings were selected to resemble a wide, medium and narrow prior, where the shrinkage effect on the standard deviations $\sigma_k^2$, $k=1,2$ caused by the inverse Gamma prior $IG(c_0,C_0)$ on $\sigma_k^2$ increases with the prior getting narrower (that is, $\sigma_i^2$ is shrunken towards zero). The same applies for the normal prior $\mathcal{N}(b_0,B_0)$ on the means $\mu_k,k=1,2$. The following hyperparameters were chosen for the three different settings: For the wide prior, $b_0:=\bar{x}$ and $B_0:=10\cdot s^2(x)$ where $\bar{x}$ and $s^2(x)$ are the complete sample mean and variance. $c_0$ and $C_0$ were selected as both $0.01$ for the wide prior, implying fatter tails of the inverse Gamma prior than in the medium or narrow prior. For the medium prior, $B_0$ was decreased to $5\cdot s^2(x)$, and $c_0$ and $C_0$ decreased to $0.1$ both. For the narrow prior finally, $B_0:=s^2(x)$ and $c_0=C_0=1$, which is the most informative of all three priors.

Subsequently, $100$ datasets with $n=100$ observations in each group were simulated, where the first group was generated as $\mathcal{N}(0,1)$ and the second as $\mathcal{N}(1,1)$. The Gibbs sampler was run for $10000$ iterations with a burn-in of $5000$ once for each prior on each dataset.
\begin{figure}[h!]
    \centering
            \includegraphics[width=0.5\textwidth]{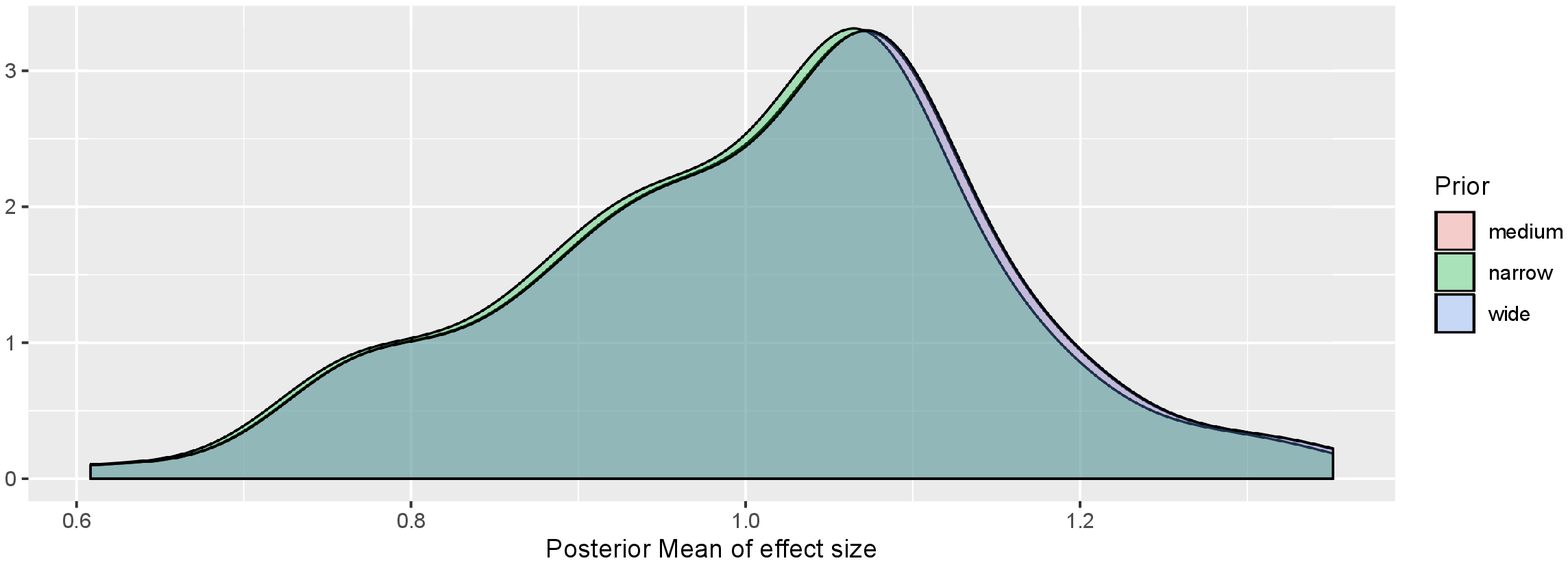}
            \includegraphics[width=0.5\textwidth]{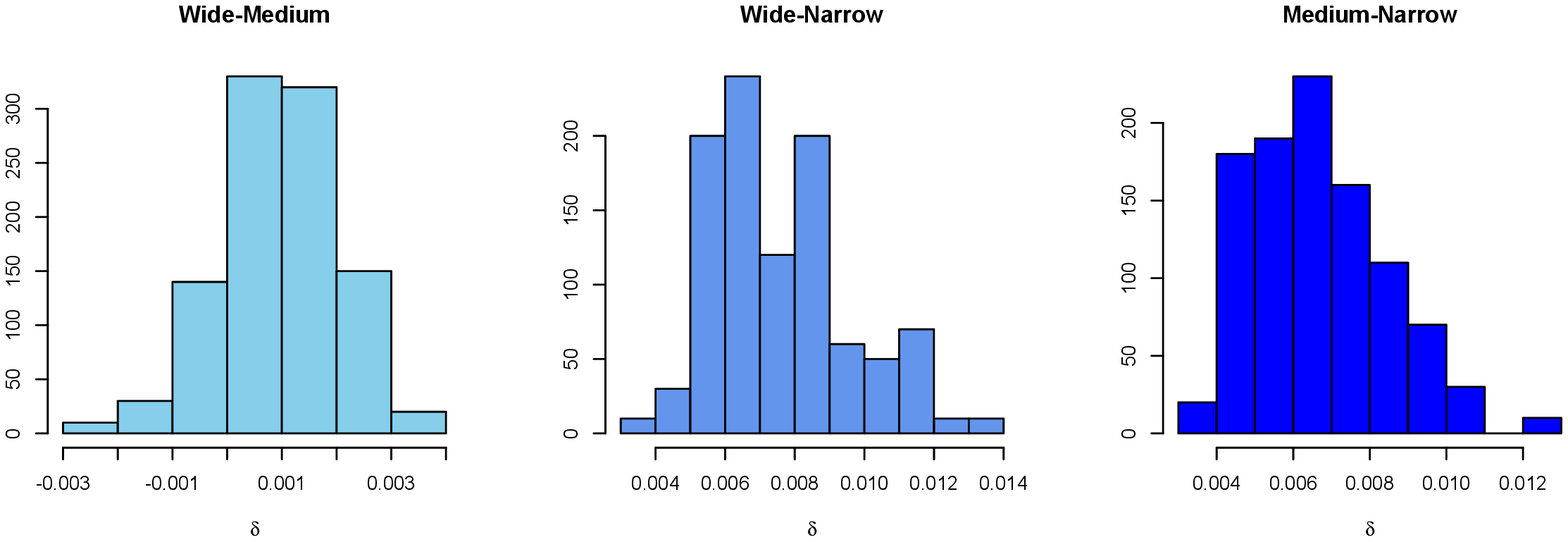}
    \caption{Prior sensitivity analysis for the $\mathcal{N}(b_0,B_0)$ prior on the means $\mu_k$ and inverse Gamma prior $IG(c_0,C_0)$ on the variances $\sigma_k^2$, $k=1,2$ for 100 datasets with first group simulated as $\mathcal{N}(0,1)$ and the second as $\mathcal{N}(1,1)$}
    \label{fig:priorSensitivity1}
\end{figure}
Fig. \ref{fig:priorSensitivity1} shows the results of the simulations. Here, the resulting posterior densities of $\delta_{MPE}$ are overlayed in figure \ref{fig:priorSensitivity1}, and it becomes clear that the wide and medium prior do result in barely differing posteriors. When using the narrow prior the shrinkage moves the posterior slightly towards smaller values of $\delta$.

The lower plot in Fig. \ref{fig:priorSensitivity1} additionally shows the posterior distributions of differences between means obtained from the three priors: The left hand plot shows the posterior distribution of differences between $\delta_{MPE}$ obtained via a wide and a medium prior. The middle plot shows the posterior distribution of differences between $\delta_{MPE}$ obtained via a wide and a narrow prior, and the right hand plot the posterior distribution of differences between $\delta_{MPE}$ obtained via a medium and a narrow prior. The results show that in all cases the differences are of tiny magnitude, indicating that the proposed t-test is quite robust to the prior hyperparameters selected. Of course, it can happen that $\delta_{MPE}$ will be drawn towards smaller values when switching from the wide to the narrow prior, but the posterior mass percentage supporting a large effect will not vary much as shown by the nearly identical resulting posterior densities in figure \ref{fig:priorSensitivity1}, which is a strength of the continuous quantification through $PMP(\delta_{MPE})$. Based on the sensitivity analysis all three hyperparameter settings differ only slightly, and therefore the wide prior seems suitable for most applications, as it places itself between the other two priors.

\section{Discussion}\label{sec:conclusionsFutureWork}

In this paper, an estimation oriented alternative to existing Bayesian t-tests was introduced. Following the proposal of a shift from hypothesis testing to estimation under uncertainty, a Bayesian two-sample t-test was derived for inference on the effect size $\delta$, which is the quantity of interest in most biomedical research. Also, the dichotomy of the ROPE decision rule of Kruschke \& Liddell \cite{Kruschke2018} was resolved by introducing the mean probable effect size $\delta_{MPE}$ as an estimator of $\delta$, combined with the posterior mass percentage $PMP(\delta_{MPE})$, a continuous measure which quantifies the support for the evidence suggested by $\delta_{MPE}$. 

Theoretical results showed that the use of the proposed method leads to a consistent estimation procedure which shifts from hypothesis testing to estimation under uncertainty. Also, theoretical results showed that under any correct ROPE $R$, the number of introduced $\alpha$ type I errors converges to zero a.s. under the law determined by the prior $\pi$ on $\delta$, while under any incorrect ROPE (which means the ROPE picked by the researcher does not cover the true parameter value $\delta_0$), the number of introduced $\alpha$ type II errors converges to zero a.s. under the law determined by the prior $\pi$ on $\delta$. Together, these properties and the introduced concept of $\alpha$-rejection make the proposed method an attractive alternative to existing solutions via p-values or the Bayes factor.

One important limitation of the approach is that the results depend on the chosen priors for $\mu_i$ and $\sigma_i$. Although quite robust, especially the choice of the inverse Gamma prior for the variances may be questioned. While it is beyond the scope of this paper to perform a sensitivity analysis using different priors for the variances, one remedy which allows changing the priors would be to switch to different sampling techniques, for example Hamiltonian Monte Carlo in Stan \cite{Carpenter2017}. Also, the speed of convergence to entire elimination of type I errors may be slow, although the simulation results are promising.

%With regard to the $p$-value in Welch's t-test, the illustrative example and the simulation study clarified the theoretical results and showed that the proposed t-test achieves better type I error control. The simulation results further indicated that sample sizes of $n=100$ in each group are large enough to reliably $\alpha$-reject a null hypothesis $H_0:\delta_0=0$ of no effect for $\alpha=.95$ when medium or large effect sizes exist. Precisely estimating the effect size required larger sample size. For small effects, sample sizes of about $n=600$ per group are necessary for precise estimation of the effect size. 

%Type I errors can be completely prevented using the proposed estimator $\delta_{MPE}$, when using sufficiently large sample sizes of about $n=200$ observations per group. 

The proposed method therefore may be helpful in improving the reproducibility in biomedical research, especially by reducing the number of false-positive results, which is one of the biggest problems of the medical sciences, see McElreath and Smaldino \cite{McElreath2015}. Finally, it should be noted that both Bayes factors as well as p-values are reasonable tools when hypothesis testing is the goal, see also the recent results of Kelter \cite{Kelter2020BayesianPosteriorIndices}. The proposed method is not intended to be a replacement of these tools, but as a complementary tool for cases in which hypothesis testing is not desired but effect size estimation under uncertainty is more important. Therefore, the main advantage may be seen in the shift towards effect size estimation. Future work could extend the model to more than two groups, leading to an equivalent of the ANOVA, use more robust component distributions like t-distributions, or derive the posterior under different priors on the mixture components parameters.

\section*{Conflict of interest}

The authors declare that they have no conflict of interest.

\section*{Appendix}

\subsection*{A.1.\enspace Derivation of the single-block Gibbs sampler}
\label{sec6}

This supplementary material provides the derivation of the joint posterior and full conditionals for the single-block Gibbs sampler.

\subsubsection*{Bayesian parameter estimation for known allocations}
In this section, for the setting when the Allocations $S$ are known the posterior distribution of $\mu_k,\sigma_k^2$ given the complete data $S,y$ are derived, see also \cite{Fruhwirth-Schnatter2006}. As already mentioned above, the weights $(\eta_1,...,\eta_K)$ are known in this case because for every observation $y_i \in (y_1,...,y_N)$ it is known to which group $y_i$ belongs, that is, the quantities $S_i=k, k\in \{1,...K\}$ are available for all $i\in \{1,...,N\}$.

In the setting of a t-test between two groups with equal sample sizes $n_1=n_2$ and $n_1+n_2=N$, the belonging of an observation $y_i$ to its group normally is known for all observations $i\in \{1,...,N\}$. The underlying data generating process therefore can be assumed to consist of a mixture of $K=2$ components with weights $\eta_1=\eta_2=0.5$. This makes inference in the mixture model much easier compared to the case when both the weights $(\eta_1,...\eta_K)$ as well as the component parameters $(\mu_1,...,\mu_K)$ and $(\sigma_1^2,...,\sigma_K^2)$ are unknown.

To conduct inference about the unknown parameters, the necessary group-specific quantities are the number $N_k(S)$ of observations in group $k$, the within-group variance $s_{y,k}^2(S)$ and the group mean $\bar{y}_k(S)$:
\begin{align*}
	&N_k(S)=|\{i:S_i=k\}|\\
	&\bar{y}_k=\frac{1}{N_k(S)}\sum_{i:S_i=k}y_i\\
	&s_{y,k}^2(S)=\frac{1}{N_k(S)}\sum_{i:S_i=k}(y_i-\bar{y}_k(S))^2	
\end{align*}
where $|\cdot|$ denotes the cardinality of a set. These quantities depend on $S$, so the classification of the observation $y_i$ to the component $S_i=k$ needs to be available. When $S_i=k$ for an observation $y_i$ holds, then the observational model for observation $y_i$ is $\mathcal{N}(\mu_k,\sigma_k^2)$ and $y_i$ contributes to the complete-data likelihood $p(y|\mu,\sigma^2,S)$ by a factor of
\begin{align*}
	\frac{1}{\sqrt{2\pi \sigma_k^2}}\text{exp}\left (-\frac{1}{2\sigma_k^2}(y_i-\mu_k)^2 \right )	
\end{align*}
Taking into account all observations $y_1,...,y_N$, the completedata likelihood function can be written as
\begin{align*}
	p(y|\mu,\sigma^2,S)=\prod_{k=1}^{K} \prod_{i:S_i=k} &(\frac{1}{2 \pi \sigma_k^2})^{N_k(S)/2}\\
	& \cdot \text{exp}(-\frac{1}{2} \sum_{i:S_i=k} \frac{(y_{i}-\mu_k)^2}{\sigma_k^2})
\end{align*}
The complete-data likelihood is a product of $K$ components, of which each summarizes the information about the $i$-th group, $i\in \{1,...,K\}$. These $K$ factors are then combined in a Bayesian analysis with a prior. While the ultimate interest lies in the posterior of both $\mu_k,\sigma_k^2$, we consider first two different cases, which will eventually lead to the solution of the joint posterior for $\mu_k$ and $\sigma_k^2$.\newline
In the first case, when the variance $\sigma_k^2$ is fixed, the complete-data likelihood function as a function of $\mu$ is the kernel of a univariate normal distribution. Choosing a $\mathcal{N}(b_0,B_0)$-distribution as a conjugate prior, the posterior density of $\mu_k$ given $\sigma_k^2$ and the $N_k(S)$ observations in group $k$ can be derived as
\begin{align}\label{eq:posterior1}
	p(\mu_k|\sigma_k^2,S,y)&\propto p(y|\mu_k,\sigma_k^2,S)\cdot p(\mu_k,\sigma_k^2,S)\\
	&\stackrel{(1)}{=}p(y|\mu_k,\sigma_k^2,S)\cdot p(\mu_k)\\
	&=(\frac{1}{2 \pi \sigma_k^2})^{N_k(S)/2}\\
	&\cdot \text{exp}(-\frac{1}{2} \sum_{i:S_i=k} \frac{(y_{i}-\mu_k)^2}{\sigma_k^2})\\
	& \cdot \frac{1}{2\pi B_0}\text{exp}(-\frac{1}{2}\frac{(\mu_k-b_0)^2}{B_0})
\end{align}
where in (1) the fact that $\sigma_k^2$ is assumed to be given and the allocations $S$ are known constants, too, was used.\newline
In general, for a sample of size $n$ from a $\mathcal{N}(\mu,\sigma)$ distribution with known variance $\sigma^2$, a standard Bayesian analysis, see e.g. \cite[p.~181]{Held2014}, yields, that the likelihood
\begin{align*}
	L(\mu)\propto \text{exp}\left ( -\frac{n}{2\sigma^2}(\mu-\bar{x})^2\right ) 	
\end{align*}
when combined with a prior $\mu \sim \mathcal{N}(\nu,\tau^2)$ leads to the posterior
\begin{align}\label{eq:finiteNormalMixturePosterior1a}
	\mu|x \sim \mathcal{N}\left ( \left ( \frac{n}{\sigma^2}+\frac{1}{\tau^2}\right )^{-1} \cdot \left ( \frac{n\bar{x}}{\sigma^2}+\frac{\nu}{\tau^2} \right ) ,\left ( \frac{n}{\sigma^2}+\frac{1}{\tau^2} \right )^{-1} \right )	
\end{align}
Substituting $\nu=b_0$ and $\tau^2=B_0$ for the prior of $\mu$ as well as $\mu_k$ for $\mu$ and $\sigma_k^2$ for $\sigma^2$ in the likelihood, the posterior $p(\mu_k|\sigma_k^2,S,y)$ in equation \ref{eq:posterior1} becomes
\begin{align}\label{eq:finiteNormalMixturePosterior2}
	&p(\mu_k|\sigma_k^2,S,y) \sim \mathcal{N}[ \left ( \frac{N_k(S)}{\sigma_k^2}+\frac{1}{B_0} \right )^{-1} \\
	&\cdot \left ( \frac{N_k(S) \bar{y}_k(S)}{\sigma_k^2}+\frac{b_0}{B_0} \right ) ,
	\left ( \frac{N_k(S)}{\sigma_k^2}+\frac{1}{B_0}  \right )^{-1}  ] \nonumber
\end{align}
By equation \ref{eq:finiteNormalMixturePosterior2}, the posterior can be written as
\begin{align*}
	\mu_k|\sigma_k^2,S,y \sim \mathcal{N}(b_k(S),B_k(S))
\end{align*}
with
\begin{align}\label{eq:finiteNormalMixturePosterior2a}
	&B_k(S)^{-1}=B_0^{-1}+\sigma_k^{-2} N_k(S)\\
	&b_k(S)=B_k(S)(\sigma_k^{-2} N_k(S) \bar{y}_k(S)+B_0^{-1} b_0)	 \label{eq:finiteNormalMixturePosterior2a2}
\end{align}
where for an empty group $k$ the term $N_k(S)\bar{y}_k(S)$ is defined as zero.\newline
On the other hand, if the mean $\mu_k$ is regarded as fixed, the complete-data likelihood as a function of $\sigma_k^2$ is the kernel of an inverse Gamma density. Choosing the conjugate inverse Gamma prior $\sigma_k^2 \sim IG(c_0,C_0)$, a standard Bayesian analysis -- for details, see for example \cite[p.~181]{Held2014} -- yields the posterior of $\sigma_k^2|\mu_k,S,y$ as
\begin{align}\label{eq:finiteNormalMixturePosterior2b}
	p(\sigma_k^2|\mu_k,S,y) \sim IG(c_k(S),C_k(S))
\end{align}
with
\begin{align}\label{eq:finiteNormalMixturePosterior2c}
	&c_k(S)=c_0+\frac{1}{2}N_k(S)\\
	&C_k(S)=C_0+\frac{1}{2}\sum_{i:S_i=k}(y_i-\mu_k)^2	 \label{eq:finiteNormalMixturePosterior2d}
\end{align}
The case of interest here is when both $\mu_k$ and $\sigma_k^2$ are unknown, and in this case a closed-form solution for the joint posterior $p(\mu_k,\sigma_k^2|S,y)$ does exist only under specific conditions. That is, the prior variance of the mean of group $k$, $\mu_k$, must depend on $\sigma_k^2$ through the relation $B_{0,k}=\frac{\sigma_k^2}{N_0}$, where $N_0$ is a newly introduced hyperparameter in the prior of $\mu_k$, that is, the prior $\mu_k \sim \mathcal{N}(b_0,B_0)$ then becomes $\mu_k \sim \mathcal{N}(b_0,\sigma_k^2 /N_0)$. The joint posterior now can be rewritten as
\begin{align}\label{eq:finiteNormalMixturePosterior3}
	&p(\mu,\sigma^2|S,y)=p(\mu_1,...,\mu_K,\sigma_1^2,...,\sigma_K^2|S,y)\\
	&\stackrel{(1)}{=}\prod_{k=1}^{K}p(\mu_k,\sigma_k^2|S,y)\stackrel{(2)}{=}\prod_{k=1}^{K}\underbrace{p(\mu_k|\sigma_k^2,S,y)}_{=:(A)}\cdot \underbrace{p(\sigma_k^2|S,y)}_{:=(B)}
\end{align}
where (1) follows from the fact that the group parameters $\mu_k,\sigma_k^2$ are assumed to be independent across groups and (2) follows from factorising the joint posterior as
\begin{align*}
		p(\mu_k,\sigma_k^2|S,y)&=\frac{p(\mu_k,\sigma_k^2|S,y)\cdot p(\sigma_k^2|S,y)}{p(\sigma_k^2|S,y)}\\
		&= p(\mu_k|\sigma_k^2,S,y)\cdot p(\sigma_k^2|S,y)
\end{align*}
As the factors $(A)$ and $(B)$ in equation \ref{eq:finiteNormalMixturePosterior3} were already derived in equation \ref{eq:finiteNormalMixturePosterior2} as well as equation \ref{eq:finiteNormalMixturePosterior2b} with corresponding parameters in equations \ref{eq:finiteNormalMixturePosterior2a}, \ref{eq:finiteNormalMixturePosterior2a2} and equations \ref{eq:finiteNormalMixturePosterior2c}, \ref{eq:finiteNormalMixturePosterior2d} for arbitrary $k$, the factor (A) of the posterior equation \ref{eq:finiteNormalMixturePosterior3} is normal-distributed $\mathcal{N}(b_k(S),B_k(S))$ with parameters
\begin{align}\label{eq:finiteNormalMixturePosterior4}
	B_k(S)&\stackrel{(1)}{=}\frac{1}{B_0^{-1}+\sigma_k^{-2}N_k(S)}\stackrel{(2)}{=}\frac{1}{\sigma_k^{-2}N_0+\sigma_k^{-2}N_k(S)}\\
	&=\frac{1}{N_0+N_k(S)}\sigma_k^2
\end{align}
and
\begin{align}
	b_k(S)&\stackrel{(3)}{=}B_k(S)(\sigma_k^{-2} N_k(S) \bar{y}_k(S)+B_0^{-1} b_0)\\
	&\stackrel{(4)}{=}B_k(S)(\sigma_k^{-2} N_k(S) \bar{y}_k(S)+\frac{N_0}{\sigma_k^2} b_0) \label{eq:finiteNormalMixturePosterior5}\\
	&\stackrel{(5)}{=}\frac{1}{N_0+N_k(S)}\sigma_k^2 (\sigma_k^{-2} N_k(S) \bar{y}_k(S)+\frac{N_0}{\sigma_k^2} b_0)\\
	&=\frac{N_k(S)\bar{y}_k(S)+N_0 b_0}{N_0+N_k(S)} \label{eq:finiteNormalMixturePosterior6}\\
	&=\frac{N_0}{N_k(S)+N_0}b_0+\frac{N_k(S)}{N_k(S)+N_0}\bar{y}_k(S)
\end{align}
where in (1) $B_k(S)^{-1}=B_0^{-1}+\sigma_k^{-2} N_k(S)$ from equation \ref{eq:finiteNormalMixturePosterior2a} was used and in (2) the relation $B_{0,k}=\frac{\sigma_k^2}{N_0}$, where $N_0$ is the newly introduced hyperparameter. In (3), equation \ref{eq:finiteNormalMixturePosterior2a2} was used, in (4) again the relation $B_{0,k}=\frac{\sigma_k^2}{N_0}$, in (5) the right-hand side of equation \ref{eq:finiteNormalMixturePosterior4} was substituted for $B_k(S)$ in equation \ref{eq:finiteNormalMixturePosterior5}.\newline
The remaining term (B) of equation \ref{eq:finiteNormalMixturePosterior3} is the marginal posterior of $\sigma_k^2$, that is
\begin{align*}
	\int p(\sigma_k^2|\mu_k,S,y)d\mu_k	
\end{align*}
and by integrating out $\mu_k$, a standard Bayesian analysis shows that the marginal posterior of $\sigma_k^2$ is distributed as inverse Gamma $IG(c_k(S),C_k(S))$, where $c_k(S)$ is already given in equation \ref{eq:finiteNormalMixturePosterior2c}, and the parameter $C_k(S)$ in equation \ref{eq:finiteNormalMixturePosterior2d} changes to
\begin{align*}
	&C_k(S)=\\
	&C_0+\frac{1}{2}\left ( N_k(S) s_{y,k}^2(S)+\frac{N_k(S) N_0}{N_k(S)+N_0}(\bar{y}_k(S)-b_0)^2 \right )	
\end{align*}
This is, because by combining an inverse-gamma prior with the normal likelihood with known mean yields an inverse-gamma posterior as shown above and marginalising this posterior for the variance yields exactly another inverse-gamma distribution with different parameters. Details can be found in \cite{Held2014}.

\subsubsection*{Application to the two-sample t-test -- Derivation of the marginal and joint posterior distributions}
In the case of the two-sample t-test, the general derivations above can be specified more precisely. For two groups, the mixture can be interpreted as a data generating process consisting of $K=2$ components. The weights $\eta_1$ and $\eta_2$ are both equal to $1/2$ for equally sized groups, that is, $N=n_1+n_2$ with $n_1$ being the sample size of group one and $n_2$ the sample size of group two and $n_1=n_2$.\newline
Taking into account all observations $y_1,...,y_N$, the complete data likelihood function can be written as
\begin{align*}
	p(y|\mu,\sigma^2,S)&=\prod_{k=1}^{2} \prod_{i:S_i=k} (\frac{1}{2 \pi \sigma_k^2})^{N_k(S)/2}\\
	& \cdot \text{exp}(-\frac{1}{2} \sum_{i:S_i=k} \frac{(y_{i}-\mu_k)^2}{\sigma_k^2})\\
	&=\prod_{i:S_i=1} (\frac{1}{2 \pi \sigma_1^2})^{N_1(S)/2} \\
	&\cdot \text{exp}(-\frac{1}{2} \sum_{i:S_i=1} \frac{(y_{i}-\mu_1)^2}{\sigma_1^2})\\
	 &\cdot \prod_{i:S_i=2} (\frac{1}{2 \pi \sigma_2^2})^{N_2(S)/2} \\
	 &\cdot \text{exp}(-\frac{1}{2} \sum_{i:S_i=2} \frac{(y_{i}-\mu_2)^2}{\sigma_2^2}) \nonumber \\
\end{align*}
where $N_1(S)=N_2(S)=N/2$.\newline
The posteriors $p(\mu_k|\sigma_k^2,S,y)$ for $k=1,2$ in equation \ref{eq:posterior1} then are $\mathcal{N}(b_k(S),B_k(S))$-distributed with
\begin{align}\label{eq:Bk}
	&B_k(S)=\frac{1}{N_0+N_k(S)}\sigma_k^2\\
	&b_k(S)=\frac{N_0}{N_k(S)+N_0}b_0+\frac{N_k(S)}{N_k(S)+N_0}\bar{y}_k(S)\label{eq:bk}
\end{align}
where also $N_1(S)=N_2(S)=N/2$ are half of total sample size and $\bar{y}_k(S)$ is the mean of group $k=1,2$. After choosing the conjugate prior $\mu_k \sim \mathcal{N}(b_0,B_0)$, these posteriors can be computed.\newline
The posteriors 
\begin{align*}
	\sigma_k^2|\mu_k,S,y \sim IG(c_k(S),C_k(S))
\end{align*}
with
\begin{align}\label{eq:ck}
	&c_k(S)=c_0+\frac{1}{2}N_k(S)\\
	&C_k(S)=C_0+\frac{1}{2}\sum_{i:S_i=k}(y_i-\mu_k)^2\label{eq:Ck}
\end{align}
for $k=1,2$ become
\begin{align*}
	&\sigma_1^2|\mu_1,S,y \sim \mathcal{G}^{-1}(c_0+\frac{1}{2}N_1(S),C_0+\frac{1}{2}\sum_{i:S_i=1}(y_i-\mu_1)^2)\\
	&\sigma_2^2|\mu_2,S,y \sim \mathcal{G}^{-1}(c_0+\frac{1}{2}N_2(S),C_0+\frac{1}{2}\sum_{i:S_i=2}(y_i-\mu_2)^2)
\end{align*}
and again, after selecting a conjugate inverse-gamma prior $\sigma_k^2 \sim IG(c_0,C_0)$ for $k=1,2$, these posteriors are also completely determined.\newline
The necessary marginal posteriors for $\sigma_k^2$ for $k=1,2$ are then obtained, following the derivations in the above section, as
\begin{align*}
	&p(\sigma_1^2|S,y) \sim IG(c_0+\frac{1}{2}N_1(S),C_0
	+\\
	&\frac{1}{2}\left ( N_1(S) s_{y,1}^2(S)+\frac{N_1(S) N_0}{N_1(S)+N_0}(\bar{y}_1(S)-b_0)^2 \right ))
\end{align*}
and
\begin{align*}
	&p(\sigma_2^2|S,y) \sim IG(c_0+\frac{1}{2}N_2(S),C_0
	\\
	&+\frac{1}{2}\left ( N_2(S) s_{y,2}^2(S)+\frac{N_2(S) N_0}{N_2(S)+N_0}(\bar{y}_2(S)-b_0)^2 \right ))
\end{align*}
These marginal posteriors are completely determined, once $c_0,C_0$ is given by the selected prior $IG(c_0,C_0)$ and the group variances $s_{y,1}^2(S)$ and $s_{y,2}^2(S)$ are calculated. Again here, $N_1(S)=N_2(S)=N/2$ due to equal sizes of both groups, and the $\bar{y}_1(S)$ and $\bar{y}_2(S)$ are the means of the two groups.\newline
The joint posterior, which is the ultimate quantity of interest, then can be rewritten as
\begin{align*}
	p(\mu,\sigma^2|S,y)&=p(\mu_1,\mu_2,\sigma_1^2,\sigma_2^2|S,y)=\prod_{k=1}^{2}p(\mu_k,\sigma_k^2|S,y)\\
	&=\prod_{k=1}^{2}\underbrace{p(\mu_k|\sigma_k^2,S,y)}_{=:(A)}\cdot \underbrace{p(\sigma_k^2|S,y)}_{:=(B)}\\
	&=p(\mu_1|\sigma_1^2,S,y)p(\sigma_1^2|S,y)\\
	&\cdot p(\mu_2|\sigma_2^2,S,y)p(\sigma_2^2|S,y)\\
	&=\mathcal{N}(b_1(S),B_1(S))\cdot IG(c_1(S),C_1(S))\\
	&\cdot \mathcal{N}(b_2(S),B_2(S))\cdot IG(c_2(S),C_2(S))
\end{align*}

\subsubsection*{A.2.\enspace Proof of Theorem \ref{theorem:fullConditionals} -- Derivation of the full conditionals for the single-block Gibbs sampler}
\begin{proof}
To make Gibbs sampling possible, the full conditionals of
\begin{align*}
	p(\mu,\sigma^2|S,y)=p(\mu_1,\mu_2,\sigma_1^2,\sigma_2^2|S,y)
\end{align*}
need to be derived. We start with $p(\mu_1|\mu_2,\sigma_1^2,\sigma_2^2,S,y)$:
\begin{align*}
	p(\mu_1&|\mu_2,\sigma_1^2,\sigma_2^2,S,y)=\frac{p(\mu_1,\mu_2,\sigma_1^2,\sigma_2^2,S,y)}{p(\mu_2,\sigma_1^2,\sigma_2^2,S,y)}\\
	&=\frac{p(\mu_1,\mu_2,\sigma_1^2,\sigma_2^2|S,y)\cancel{p(S,y)}}{p(\mu_2,\sigma_1^2,\sigma_2^2|S,y)\cancel{p(S,y)}}\\
	&\stackrel{(1)}{=}\frac{\prod_{k=1}^{K}p(\mu_k|\sigma_k^2,S,y)p(\sigma_k^2|S,y)}{p(\mu_2,\sigma_2^2|S,y)p(\sigma_1^2|S,y)}\\
	&\stackrel{(2)}{=}\frac{p(\mu_1|\sigma_1^2,S,y)\cancel{p(\sigma_1^2|S,y)}\cancel{p(\mu_2|\sigma_2^2,S,y)}\cancel{p(\sigma_2^2|S,y)}}{\cancel{p(\mu_2|\sigma_2^2,S,y)}\cancel{p(\sigma_2^2|S,y)}\cancel{p(\sigma_1^2|S,y)}}\\
	&=p(\mu_1|\sigma_1^2,S,y)
\end{align*}
where in (1) the independence of $\sigma_1^2$ and $\mu_2,\sigma_2^2$ was used, so that $p(\mu_2,\sigma_1^2,\sigma_2^2|S,y)=p(\mu_2,\sigma_2^2|S,y)\cdot p(\sigma_1^2|S,y)$ holds, and in (2) the factorization $p(\mu_2,\sigma_2^2|S,y)=p(\mu_2|\sigma_2^2,S,y)\cdot p(\sigma_2^2|S,y)$ was used.

The full conditional of $\mu_2$ is given by $p(\mu_2|\mu_1,\sigma_1^2,\sigma_2^2,S,y)$, which is derived as
\begin{align*}
	p(\mu_2|\mu_1,&\sigma_1^2,\sigma_2^2,S,y)=\frac{p(\mu_2,\mu_1,\sigma_1^2,\sigma_2^2,S,y)}{p(\mu_1,\sigma_1^2,\sigma_2^2,S,y)}\\
	&=\frac{p(\mu_2,\mu_1,\sigma_1^2,\sigma_2^2|S,y)\cancel{p(S,y)}}{p(\mu_1,\sigma_1^2,\sigma_2^2|S,y)\cancel{p(S,y)}}\\
	&=\frac{\prod_{k=1}^{K}p(\mu_k|\sigma_k^2,S,y)p(\sigma_k^2|S,y)}{p(\mu_1,\sigma_1^2|S,y)p(\sigma_2^2|S,y)}\\
	&=\frac{\cancel{p(\mu_1|\sigma_1^2,S,y)}\cancel{p(\sigma_1^2|S,y)}p(\mu_2|\sigma_2^2,S,y)\cancel{p(\sigma_2^2|S,y)}}{\cancel{p(\mu_1|\sigma_1^2,S,y)}\cancel{p(\sigma_2^2|S,y)}\cancel{p(\sigma_1^2|S,y)}}\\
	&=p(\mu_2|\sigma_2^2,S,y)
\end{align*}
where the reasoning is the same as in the derivation of the full conditional for $\mu_1$.

The full conditional of $\sigma_1^2$ is $p(\sigma_1^2|\mu_1,\mu_2,\sigma_2^2,S,y)$, which is derived as:
\begin{align*}
	p(&\sigma_1^2|\mu_1,\mu_2,\sigma_2^2,S,y)=\frac{p(\sigma_1^2,\mu_1,\mu_2,\sigma_2^2,S,y)}{p(\mu_1,\mu_2,\sigma_2^2,S,y)}\\
	&=\frac{p(\sigma_1^2,\mu_1,\mu_2,\sigma_2^2|S,y)\cancel{p(S,y)}}{p(\mu_1,\mu_2,\sigma_2^2|S,y)\cancel{p(S,y)}}\\
	&=\frac{p(\sigma_1^2,\mu_1,\mu_2,\sigma_2^2|S,y)}{p(\mu_1,\mu_2,\sigma_2^2|S,y)}\\
	&\stackrel{(1)}{=}\frac{\prod_{k=1}^{2}p(\sigma_k^2|\mu_k,S,y)\cdot p(\mu_k|S,y)}{p(\mu_1|S,y)\cdot p(\mu_2,\sigma_2^2|S,y)}\\
	&\stackrel{(2)}{=}\frac{p(\sigma_1^2|\mu_1,S,y)\cdot p(\mu_1|S,y)\cdot \cancel{p(\sigma_2^2|\mu_2,S,y)}\cdot \cancel{p(\mu_2|S,y)}}{p(\mu_1|S,y)\cdot \cancel{p(\sigma_2^2|\mu_2,S,y)}\cdot \cancel{p(\mu_2|S,y)}}\\
	&=\frac{\cancel{p(\mu_1|S,y)}\cdot p(\sigma_1^2|\mu_1,S,y)}{\cancel{p(\mu_1|S,y)}}\\
	&=p(\sigma_1^2|\mu_1,S,y)
\end{align*}
where in (1) first the independence of parameters between both groups was used, that is, the independence of $\mu_1,\sigma_1^2$ and $\mu_2,\sigma_2^2$ and second (as a special case of this fact) the independence of $\mu_1$ and $\mu_2,\sigma_2^2$ was used. Therefore,  $p(\sigma_1^2,\mu_1,\mu_2,\sigma_2^2|S,y)=\prod_{k=1}^{2}p(\sigma_k^2|\mu_k,S,y)\cdot p(\mu_k|S,y)$ holds and also $p(\mu_1,\mu_2,\sigma_2^2|S,y)=p(\mu_1|S,y)\cdot p(\mu_2,\sigma_2^2|S,y)$. In (2) the factorization $p(\mu_2,\sigma_2^2|S,y)=p(\sigma_2^2|\mu_2,S,y)\cdot p(\mu_2|S,y)$ was used.

The full conditional $p(\sigma_2^2|\mu_1,\mu_2,\sigma_1^2,S,y)$ of $\sigma_2^2$ is similarly given by:
\begin{align*}
	p(&\sigma_2^2|\mu_1,\mu_2,\sigma_1^2,S,y)=\frac{p(\sigma_1^2,\mu_1,\mu_2,\sigma_2^2,S,y)}{p(\mu_1,\mu_2,\sigma_1^2,S,y)}\\
	&=\frac{p(\sigma_1^2,\mu_1,\mu_2,\sigma_2^2|S,y)\cancel{p(S,y)}}{p(\mu_1,\mu_2,\sigma_1^2|S,y)\cancel{p(S,y)}}\\
	&=\frac{p(\sigma_1^2,\mu_1,\mu_2,\sigma_2^2|S,y)}{p(\mu_1,\mu_2,\sigma_1^2|S,y)}\\
	&\stackrel{(1)}{=}\frac{\prod_{k=1}^{2}p(\sigma_k^2|\mu_k,S,y)\cdot p(\mu_k|S,y)}{p(\mu_2|S,y)\cdot p(\mu_1,\sigma_1^2|S,y)}\\
	&\stackrel{(2)}{=}\frac{\cancel{p(\sigma_1^2|\mu_1,S,y)}\cdot \cancel{p(\mu_1|S,y)}\cdot p(\sigma_2^2|\mu_2,S,y)\cdot p(\mu_2|S,y)}{p(\mu_2|S,y)\cdot \cancel{p(\sigma_1^2|\mu_1,S,y)}\cdot \cancel{p(\mu_1|S,y)}}\\
	&=\frac{\cancel{p(\mu_2|S,y)}\cdot p(\sigma_2^2|\mu_2,S,y)}{\cancel{p(\mu_2|S,y)}}\\
	&=p(\sigma_2^2|\mu_1,S,y)
\end{align*}
where in (1) first the independence of parameters between both groups, that is, the independence of $\mu_1,\sigma_1^2$ and $\mu_2,\sigma_2^2$ and second (as a special case of this) the independence of $\mu_2$ and $\mu_1,\sigma_1^2$ was used. Therefore, $p(\sigma_1^2,\mu_1,\mu_2,\sigma_2^2|S,y)=\prod_{k=1}^{K}p(\sigma_k^2|\mu_k,S,y)\cdot p(\mu_k|S,y)$ holds and one also has $p(\mu_1,\mu_2,\sigma_2^2|S,y)=p(\mu_1|S,y)\cdot p(\mu_2,\sigma_2^2|S,y)$. In (2) the factorization $p(\mu_1,\sigma_1^2|S,y)=p(\sigma_1^2|\mu_1,S,y)\cdot p(\mu_1|S,y)$ was used. In total, the full conditionals therefore are derived as:
\begin{align*}
		&p(\mu_1|\mu_2,\sigma_1^2,\sigma_2^2,S,y)=p(\mu_1|\sigma_1^2,S,y)\\
		&p(\mu_2|\mu_1,\sigma_1^2,\sigma_2^2,S,y)=p(\mu_2|\sigma_2^2,S,y)\\
		&p(\sigma_1^2|\mu_1,\mu_2,\sigma_2^2,S,y)=p(\sigma_1^2|\mu_1,S,y)\\
		&p(\sigma_2^2|\mu_1,\mu_2,\sigma_1^2,S,y)=p(\sigma_1^2|\mu_1,S,y)
\end{align*}
When using the independence prior, the full conditionals are therefore given by
\begin{align}\label{eq:fullConditionals1}
		&p(\mu_1|\mu_2,\sigma_1^2,\sigma_2^2,S,y)=p(\mu_1|\sigma_1^2,S,y)\sim \mathcal{N}(b_1(S),B_1(S))\\
		&p(\mu_2|\mu_1,\sigma_1^2,\sigma_2^2,S,y)=p(\mu_2|\sigma_2^2,S,y)\sim \mathcal{N}(b_2(S),B_2(S))\\
		&p(\sigma_1^2|\mu_1,\mu_2,\sigma_2^2,S,y)=p(\sigma_1^2|\mu_1,S,y)\sim IG(c_1(S),C_1(S))\\
		\label{eq:fullConditionals4}
&p(\sigma_2^2|\mu_1,\mu_2,\sigma_1^2,S,y)=p(\sigma_1^2|\mu_2,S,y)\sim IG(c_2(S),C_2(S))\end{align}
with $b_1(S),B_1(S),b_2(S),B_2(S),c_1(S),c_2(S),C_1(S)$ and $C_2(S)$ as specified in the previous section in equations (\ref{eq:Bk}), (\ref{eq:bk}), (\ref{eq:ck}) and (\ref{eq:Ck}), which completes the proof.
\end{proof}

\subsubsection*{A.3.\enspace Proof of Corollary \ref{corollary:gibbsSampler}}
\begin{proof}
From standard MCMC theory, see e.g. \cite{Robert2004}, it is clear that the full conditionals derived in Theorem \ref{theorem:fullConditionals} can be used to construct a Gibbs sampler, by iteratively updating each parameter via simulating step by step from the full conditionals (\ref{eq:fullConditionals1}) to (\ref{eq:fullConditionals4}). Using (\ref{eq:fullConditionals1}) to (\ref{eq:fullConditionals4}), this leads to the following Gibbs sampling algorithm:
\begin{enumerate}
	\item{Sample $\sigma_k^2$ in each group $k$, $k=1,2$ from an inverse Gamma distribution
		$\mathcal{G}^{-1}(c_k(S),C_k(S))$ (which depends on $\mu_k$)}
	\item{Sample $\mu_k$ in each group $k$, $k=1,2$, from a normal distribution $\mathcal{N}(b_k(S),B_k(S))$ (which depends on $\sigma_k^2$)}
\end{enumerate}
where $B_k(S), b_k(S)$ and $c_k(S), C_k(S)$ are given by equations (\ref{eq:Bk}), (\ref{eq:bk}), (\ref{eq:ck}) and (\ref{eq:Ck}). The convergence to the joint posterior $p(\mu_1,\mu_2,\sigma_1^2,\sigma_2^2|S,y)$ then follows then from standard MCMC theory, see \cite{Robert2004}.
\end{proof}

\subsection*{A.4.\enspace Proof of Theorem \ref{theorem:consistencyDeltaMPE}}
\begin{proof}
	 	In the case the ROPE $R_j$ is correct, $R_j$ includes the true effect size $\delta_0$, so that $\delta_0 \subset R_j$. Estimation of $\delta$ via $\delta_{MPE}$ is then consistent: A posterior distribution $p_n$ is said to be consistent for the parameter $\delta_0$, if for every neighbourhood $U$ of $\delta_0$, $p_n(U)\xrightarrow[n \rightarrow \infty ]{a.s.}1$ almost surely under the law determined by $\delta_0$. By Theorem 1 in \cite{ghosal1996} the consistency of the posterior follows for any prior $\pi$ when choosing \textit{any} ROPE $U$ including the true parameter $\delta_0$, except possibly on a set of $\pi$-measure zero. As a direct consequence one therefore obtains:
		If the ROPE $R_j \neq \emptyset$ is correct and contains the true parameter $\delta_0$, any prior $\pi$ leads to a consistent posterior for which $p_n(R_j)\xrightarrow[n \rightarrow \infty ]{a.s.}1$ almost surely under the law determined by $\delta_0$ except possibly on a set of $\pi$-measure zero. This means that
		\begin{align*}
			PMP(\delta_{MPE})=\int_{R_j} f(\theta|x)d\theta \xrightarrow[n \rightarrow \infty ]{a.s.}1
		\end{align*} 
		If on the other hand $R_j$ is incorrect, then $\delta_0 \notin R_j$. Then there exists a neighbourhood $N:=(\delta_0-\varepsilon,\delta_0+\varepsilon)$ for $\varepsilon >0$ around $\delta_0$, so that $N\cap R_j=\emptyset$. Then, as $p_n(N)\xrightarrow[n \rightarrow \infty ]{a.s.}1$ almost surely under the law determined by $\delta_0$ except possibly on a set of $\pi$-measure zero, it follows that on the complement $N^c$, $p_n(N^c)\xrightarrow[n \rightarrow \infty ]{a.s.}0$ and because of $R_j \subset N^c$ also that $p_n(R_j)\xrightarrow[n \rightarrow \infty ]{a.s.}0$ almost surely under the law determined by $\delta_0$ except possibly on a set of $\pi$-measure zero. Thereby it follows that
		\begin{align*}
			PMP(\delta_{MPE})=\int_{R_j} f(\theta|x)d\theta \xrightarrow[n \rightarrow \infty ]{a.s.}0
		\end{align*}
		
%		The decision rule based on the ROPE is thus consistent, and therefore the decision rule based on $\delta_{MPE}$ is, too, as $PMP(\delta_{MPE})\rightarrow 1$ for $n\rightarrow \infty$ as all posterior probability mass will be allocated inside the ROPE including $\theta_0$ for $n\rightarrow \infty$.
	 \end{proof}
	 
\subsection*{A.5.\enspace Proof of Theorem \ref{theorem:errors}}
\begin{proof}
	An $\alpha$ type I error happens if the true parameter value $\delta_0 \in H$, with $H\subset R$ for a ROPE $R\subset \Theta$, but $H$ is $\alpha$-rejected for $\alpha$. If any correct ROPE $R$ is selected around the hypothesis $H\subset \Theta$ which makes a statement about the unknown parameter $\delta$, then the true value $\delta_0$ of $\delta$ is inside $R$, that is $\delta_0 \subset R$. Then, under any prior $\pi$ on $\delta$, the posterior $p_n(R)\xrightarrow[n \rightarrow \infty ]{a.s.}1$ except on a set of $\pi$-measure zero, compare \cite{ghosal1996}. Therefore, the corresponding $\alpha$\% HPD interval $C_{\alpha}$ of $f(\delta|x)$ lies inside $R$ for $n\rightarrow \infty$, too, that is: $C_{\alpha} \subset R$, and by definition, $H$ is then $\alpha$-accepted. As $\alpha$ was arbitrary, the above holds in particular without loss of generality for $\alpha=1$, and therefore, $H$ is accepted always for $n\rightarrow \infty$ almost surely under the law determined by $\pi$ for any correct ROPE $R$ around the hypothesis $H$. This in turn implies that $H$ can only be $\alpha$-rejected for $\alpha=0$ under the above conditions.\footnote{Note that $\alpha$ rejection of $H$ for $\alpha=0$ is only stating that zero percent of the HPD interval are located outside the ROPE $R$ around $H$, which means that the $\alpha$\% HPD lies fully inside the ROPE. This in fact confirms $H$, being in agreement with the intuition of no type I error occuring.}
	
	An $\alpha$ type II error happens if the true parameter value $\delta_0 \notin H$ and $\delta \notin R$, with $H\subset R$ for a ROPE $R\subset \Theta$, but $H$ is $\alpha$-accepted for $\alpha$. If any incorrect ROPE $R$ is selected around the hypothesis $H\subset \Theta$ which makes a statement about the unknown parameter $\delta$, then the true value $\delta_0$ of $\delta$ is not inside $R$, that is $\delta_0 \notin R$. Then, under any prior $\pi$ on $\delta$, the posterior $p_n(R)\xrightarrow[n \rightarrow \infty ]{a.s.}0$ except on a set of $\pi$-measure zero, compare \cite{ghosal1996}. Therefore, the corresponding $\alpha$\% HPD interval $C_{\alpha}$ of $f(\delta|x)$ lies not inside $R$ for $n\rightarrow \infty$, which means $C_{\alpha} \notin R$, and by definition, $H$ is then $\alpha$-rejected. As $\alpha$ was arbitrary, the above holds in particular for $\alpha=1$, and therefore, $H$ is rejected always for $n\rightarrow \infty$ almost surely under the law determined by $\pi$ for any incorrect ROPE $R$ around the hypothesis $H$. This also implies that $H$ can only be $\alpha$-accepted for $\alpha=0$ under the above conditions.\footnote{Here too, $\alpha$-acceptance for $\alpha=0$ can be interpreted as zero percent of the HPD being located inside the ROPE R around $H$, being in agreement with the intuition of no type II error occuring.}
\end{proof}

\section*{Compliance with Ethical Standards}
\section*{Funding}
Not applicable.
\section*{Conflicts of interest}
The author declares no conflict of interest.
\section*{Availability of data and material}
All data used in the simulations and illustrative example are available at \url{https://osf.io/cvwr5/}.
\section*{Code availability}
A complete replication script to run the simulations and produce all results is available at the Open Science Framework under \url{https://osf.io/cvwr5/}. The proposed Bayesian t-test is made available in the R package \texttt{bayest} at CRAN: \url{https://cran.r-project.org/web}\newline\url{/packages/bayest/index.html}.
%\begin{acknowledgements}
%\end{acknowledgements}


\bibliographystyle{spmpsci}      % mathematics and physical sciences
\bibliography{library}   % name your BibTeX data base

\end{document}